\documentstyle[prl,aps,multicol,psfig,epsfig]{revtex}
\begin{document}
\draft
\widetext
\title{ Microscopic theory of weak pseudogap behavior  
in the underdoped cuprate  \\
 superconductors  I:  General theory and  quasiparticle properties }
\author{J\"org Schmalian,$^{a}$ David Pines,$^{a,b}$
 and   Branko Stojkovi\'c$^{b}$    }
\address{$^{a}$University of Illinois at 
Urbana-Champaign,   Loomis Laboratory 
of Physics, 1110 W. Green,  Urbana, IL, 61801\\
$^{b}$Center for Nonlinear Studies, Los Alamos National Laboratory, 
Los Alamos, NM, 87545 }
\date{\today}
\maketitle
\widetext 
\begin{abstract} 
\leftskip 54.8pt
\rightskip 54.8pt 
We use a  novel solution of the spin fermion model  which is valid 
in the quasi-static limit $\pi T \gg \omega_{\rm sf}$,   found
  in the intermediate (pseudoscaling) regime  of 
the magnetic phase diagram of cuprate superconductors,  to obtain 
results for the   temperature and doping dependence of
the single particle spectral density, the electron-spin fluctuation
vertex function, and the low frequency dynamical spin susceptibility.
The resulting strong anisotropy of the  spectral density and the
vertex function
  lead to the qualitatively different behavior of {\em hot}  
(around ${\bf k}=(\pi,0)$) and {\em cold} (around ${\bf
  k}=(\pi/2,\pi/2)$) 
quasiparticles seen in ARPES experiments.
 We find  that  the broad  high energy  features
found in ARPES measurements  of the  spectral density of the
 underdoped cuprate superconductors are
determined by  strong  antiferromagnetic (AF) 
correlations and incoherent  precursor effects
of an SDW state, with reduced renormalized effective coupling
constant.
Due to this transfer of spectral weight to higher energies, the
low frequency spectral weight of {\em hot} states 
is strongly reduced  but couples very strongly to the spin excitations 
of the system.
For realistic values of  the antiferromagnetic correlation  
length, their Fermi surface changes  its general  shape only slightly
but the strong scattering  of hot states makes the Fermi surface crossing
 invisible above a pseudogap temperature  $T_*$.
 The electron spin-fluctuation vertex function,
i.e. the   effective interaction of low energy
quasiparticles
and spin degrees of freedom,  is found to be 
strongly anisotropic and  enhanced for
hot quasiparticles; the corresponding
 charge-fluctuation vertex is considerably  diminished.
We thus 
 demonstrate
that, once established, strong AF correlations act to reduce
substantially the effective electron-phonon coupling constant in
cuprate superconductors.  
 \end{abstract}   
\begin{multicols}{2}
\narrowtext   
\section{Introduction}

        In addition to their high transition temperatures and the d$_{x^2-y^2}$
symmetry of their superconducting state, the cuprate superconductors
possess a remarkable range of normal state anomalies. Seen first as charge
response anomalies in transport, Raman, and optical experiments, and
subsequently as spin response anomalies in
nuclear magnetic resonance (NMR) and inelastic neutron scattering 
(INS)  experiments, recent
specific heat and angular resolved photoemission spectroscopy
(ARPES) 
experiments have shown that these anomalies are accompanied by, and may
indeed originate in, anomalous planar quasiparticle behavior. It is
convenient to discuss the temperature and doping dependence of this
"uniformly" anomalous behavior in terms of the schematic phase diagram
shown in Fig. 1. There one sees that overdoped and underdoped systems may
be distinguished by the extent to which these exhibit   crossover
behavior in the normal state:  underdoped systems 
 exhibit two distinct crossovers in normal state behavior before
going superconducting, while overdoped systems pass directly from a single
class of anomalous normal state behavior to the superconducting 
state~\cite{general}. In a broader perspective of 
this schematic phase diagram, which is applicable to the
 YBa$_2$Cu$_3$O$_{7-\delta}$,
YBa$_2$Cu$_4$O$_{8}$,  
Bi$_2$Sr$_2$Ca$_{1}$Cu$_2$O$_{8+\delta}$,
 HgBa$_2$CuO$_{4+\delta}$, HgBa$_2$Ca$_2$Cu$_3$O$_8$
and Tl$_2$Ba$_2$Ca$_2$Cu$_3$O$_{10}$ 
  systems, if one defines an optimally doped system as that which
possesses the highest superconducting transition temperature within a given
family, then optimally doped systems are in fact underdoped.

Attempts to understand the different regimes of this phase diagram have been
based on strong magnetic
 precursors~\cite{TT1_a,TT1_b,TT1_c,TT1_e,TT1_f ,SGB97}, 
the formation of
dynamical charge modulations in form of stripes~\cite{TT2}, the
appearance of preformed Cooper pairs above
 $T_{\rm c}$~\cite{EKiv95,ML96,ERS97} , or the
separation of spin and charge degrees of
 freedom~\cite{TT3_a,TT3_b}.

 A phase diagram similar to Fig. 1 was independently derived from
studies of the charge response by Hwang, Batlogg, and 
their collaborators~\cite{Batlogg}
and from an analysis of the low frequency NMR
experiments~\cite{NMRreview} 
by Barzykin and
Pines~\cite{BP95}. The latter authors identified 
the upper crossover temperature,
T$^{\rm cr}$,
from measurements of the uniform susceptibility, $\chi _o$ , in Knight shift
experiments, which show that for underdoped systems $\chi _o$  possesses a
maximum at a temperature T$^{\rm cr}$, which in underdoped systems, increases rapidly
from T$_{\rm c}$ as the doping level is reduced. The fall-off in susceptibility for
temperatures below T$^{\rm cr}$ was first studied in 
detail by Alloul {\em et al.}~\cite{AMCM88}, and led
Friedel~\cite{Fri89}  to propose that it might arise from a
 near spin density wave (SDW) instability; he
coined the term pseudogap to explain its behavior, in analogy to the
quasiparticle pseudogap seen in charge density wave (CDW)
 systems.
Barzykin and Pines identified
further crossover behavior in this pseudogap regime by examining the
behavior of the $^{63}$Cu nuclear spin-lattice relaxation
time, $^{63}T_1$ , as the
temperature was reduced below T$^{\rm cr}$. 
They noted that between T$^{\rm cr}$ and a lower
crossover temperature, T$^*$, the product, $^{63}T_1 T$  decreases linearly in
temperature, while shortly below T$^*$ this product has a minimum, followed by
an increase as the temperature is further lowered, an increase which is
strongly suggestive of gap-like behavior. They proposed that these two
crossovers were accompanied by changes in dynamical scaling behavior which
could be measured directly if NMR measurements of $^{63}T_1$   could be accompanied
by measurements of the spin-echo decay time, $^{63}T_{2G} $.  Above T$^{\rm cr}$ they argued
that the ratio, $^{63}T_1T /^{63}T_{2G}^2$,   would be independent of temperature, a result
equivalent to arguing that the characteristic energy of the spin
fluctuations, $\omega_{\rm sf}$ , would be proportional to the inverse square of the
antiferromagnetic correlation length, $\xi$. Between T$^{\rm cr}$ and T$^*$ they proposed
that the ratio, $^{63}T_1 T /^{63}T_{2G}$    would be independent of temperature,which means
that an underdoped system would exhibit $z=1$ scaling behavior, i.e.   $\omega_{\rm sf}$
would be proportional to $\xi^{-1}$ ;  below T$^*$ they found that
the increase in
$\omega_{\rm sf}$  would be accompanied by a freezing out of the
temperature-dependent antiferromagnetic correlations; i.e. $\xi^{-1}$, which was
proportional to $a + bT$ between T$^{\rm cr}$ and T$^*$, would approach a constant.This
behavior has recently been confirmed in NMR measurements on YBa$_2$Cu$_4$O$_8$ by
Curro {\em et al.}~\cite{CCS96} while $z=1$
  pseudoscaling behavior has been found in INS experiments
on La$_{1.86}$Sr$_{0.14}$Cu O$_4$ by Aeppli  {\em et al.}~\cite{AMH97}.

        Because pseudogap behavior is found both between T$^{\rm cr}$ and T$^*$, and
between T$^*$ and T$_{\rm c}$, the terms weak pseudogap and strong pseudogap behavior
were coined to distinguish between the two 
regimes~\cite{SPS97}. Thus in the weak
pseudogap regime one finds $z=1$ pseudoscaling  (because the scaling behavior
is not universal) behavior, with both $\omega_{\rm sf}$  and $\xi^{-1}$  exhibiting linear in
$T$ behavior, while the rapid increase in $^{63}T_1T$, or what is
equivalent,   $\omega_{\rm sf}$, found below T$^*$ suggests that
strong pseudogap
 is an appropriate  descriptor for this behavior.

        An alternative perspective on weak and strong pseudogap behavior
comes from ARPES~\cite{ZX95,Ding} 
 and tunneling experiments~\cite{RF97} , which focus directly on
 single
particle excitations. Above T$^*$, ARPES experiments show that the spectral
density of quasiparticles located near the $(\pi,0)$   part of the Brillouin zone,
develops a high energy feature, a result which suggests that the transfer of
spectral weight from low energies to high energies for part of the
quasiparticle spectrum may be the physical origin of the weak pseudogap
behavior seen in NMR experiments. Below T$^*$, ARPES experiments
disclose the presence of  a
leading-edge gap, a momentum-dependent shift of the lowest binding energy
relative to the chemical potential by an amount up to   $30\, {\rm meV}$ for
quasiparticles near $(\pi,0)$;    it seems natural to associate
the strong pseudogap behavior seen in the NMR experiments with this leading
edge gap. Recent tunneling experiments have shown that both the high energy
feature (i.e. weak pseudogap behavior) and the strong single particle
pseudogap can also be observed in the tunneling conductance, with the high
energy feature occurring primarily in the  occupied part of the spectrum.
 
Strong pseudogap behavior is also
 seen in specific heat, d.c.   transport,   optical experiments, 
and Raman experiments.
Below $T_*$, a reduced scattering rate   for frequencies
$\omega < \pi T_*$ has been extracted from the
optical conductivity using a single band picture~\cite{PBT96}.
 This  suggests
that excitations  in      the pseudogap regime
are more coherent then expected by   extrapolation 
from higher temperatures. This point of view is 
supported by recent Raman 
experiments~\cite{BKK97,NOH97} which observe 
in the B$_{1g}$-channel, sensitive to single particle
states around ${\bf k}=(\pi,0)$, a suppression of the broad incoherent
 Raman  continuum
and a rather sharp structure at about twice the single
particle gap of ARPES experiments~\cite{BKK97}.
Resistivity measurements also show that  below  $T_*$ the systems gets
more conducting than one would have expected from   the 
linear    resistivity  at   higher  temperatures~\cite{res2}. 

\begin{figure}
\centerline{\epsfig{file=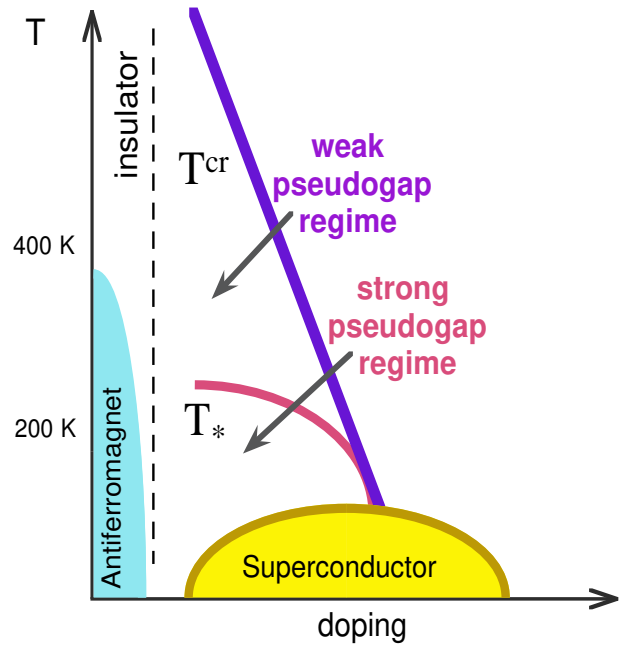,width=10cm, height=10cm,scale=1}}
\caption{Schematic phase diagram of underdoped cuprates. Note the
  presence of two different crossover temperatures:
  $T^{\rm cr}$, which characterizes the onset of sizable
  antiferromagnetic correlations; and $T^*$, which  signals the onset 
of a considerable loss of low energy spectral weight in the
quasiparticle
spectrum leading to a minimum of the characteristic spin-fluctuation
energy
$\omega_{\rm sf}$. The region between  $T^{\rm cr}$ and $T^*$  is the
weak pseudogap regime discussed in this paper. }
\label{fI1}
\end{figure}

It  is natural  to believe that  the pseudogap in the spin damping, as 
observed in $^{63}T_1T$ measurements,  
the single particle pseudogap of ARPES and tunneling experiments, and
the  pseudogap of the scattering rate  are
closely related   and must be understood simultaneously.
 Furthermore,    it is essential  for any  theory 
of the strong pseudogap  to account properly 
for the already existent anomalies above  $T_*$, 
because they  are likely caused by the same 
 underlying  effective
interactions. As can be seen by inspection  of  Fig.~\ref{fI1},
strong pseudogap behavior in underdoped cuprates
only occurs once the system has passed the 
weak pseudogap state.

In this paper we  concentrate on the weak pseudogap (pseudoscaling) regime
above $T_*$  and give a more detailed account of the  preliminary
   results we obtained using a spin-fluctuation model of normal state
behavior~\cite{SPS97}. 
It is our  aim to provide a  
 quantitative understanding for $T>T_*$  of quantities
reflecting  strong pseudogap behavior  below $T_*$.
We derive   a novel solution of the  spin fermion model in the 
quasi-static limit $\pi T \gg \omega_{\rm sf}$,  
 relevant   for the intermediate weak pseudogap  regime  of 
the phase diagram.
We demonstrate that  the broad  high energy  features of the  spectral density
found in ARPES measurements    of 
underdoped cuprates   are
determined by  strong  antiferromagnetic 
correlations and incoherent  precursor effects
of an SDW state.
The spectral density at the Fermi energy  and the electron-spin fluctuation
vertex function    are strongly anisotropic,
leading to qualitatively different behavior of {\em hot}  
(around ${\bf k}=(\pi,0)$) and {\em cold} (around ${\bf k}=(\pi/2,\pi/2)$) 
momentum states, whereas the
Fermi surface itself   changes    only slightly.  
We present new 
results for the effective interaction of quasiparticles with spin and charge collective
modes.
In distinction to the strong coupling of hot quasiparticles to spin
excitations, we demonstrate that their renormalized coupling to
charge degrees of freedom, including phonons, is strongly suppressed.
Finally, we show that the onset temperature, $T^{\rm cr}$, 
of weak pseudogap    behavior is determined by 
the strength, $\xi$, of the 
AF correlations.

Our theory also allows us to investigate the low frequency spin
and charge response functions. In a subsequent publication, we will
discuss the  suppression of the spin damping and
further  generic changes in low frequency magnetic 
behavior seen in NMR experiments as well as   the optical response,
particularly as far as  the B$_{1g}$-Raman continuum is concerned.

The paper is organized as follows. In the next section we summarize
important findings of ARPES results which will later on be explained
 by  our  theory of the weak pseudogap
regime. Next, we give 
the basic concept of the spin fluctuation   model  and derive the spin
fermion  model. In the following,  fourth, section we discuss in detail 
our   solution of the spin fermion model in the quasi-static limit,
with particular attention to  the new physics of the spin fermion model for  intermediate coupling.
Our solution is obtained by the complete summation of the perturbation series,
 and is motivated in part by  a  theory for one dimensional charge density wave
 systems developed  by Sadovskii~\cite{Sad79}. We have extended his theory to
 the case of two spatial dimensions and isotropic spin fluctuations
 and, in so doing,
found that we  could
avoid several technical problems of the earlier approach.
Technical details of the   rules we used for computing diagrams are presented in   
Appendix~\ref{App_GF} and B.
Readers not interested in these technical aspects  can skip the theory section and 
should   be able to follow   the discussion of our  results
for the spectral density and vertex functions  in 
 the  fifth   and sixth sections, respectively.
In particular, 
 results for the single particle properties are discussed 
at length and compared
 with ARPES  experiments. Finally our theory for the weak pseudogap regime
is summarized in the last section, where we also  consider 
the physics of the strong pseudogap state and summarize some
predictions and consequences of our theory.
We argue that a proper description of the higher temperature  weak pseudogap
regime is essential for a further investigation 
of the low temperature  strong pseudogap state and argue
that the strong pseudogap state and precursors in the pairing channel
are the quantum manifestation of strong
antiferromagnetic correlations whereas the spin density wave
precursors are the classical 
manifestation of it.

\section{ARPES experiments}

ARPES experiments offer a powerful probe   of the
quasiparticle properties  of cuprates.
Since they   provide  unusually strong experimental 
 constraints for any theory of optimally doped and underdoped
cuprate superconductors, 
we summarize in this section  the main experimental
results obtained  by  this experimental technique  

\begin{figure}
\centerline{\epsfig{file=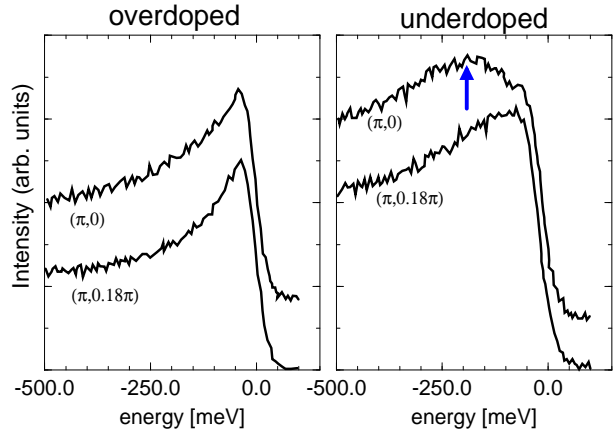,width=10cm, height=8.2cm, scale=1}}
\caption{ ARPES spectra  from Ref.[30] for  momenta  ${\bf  k}$ at and  
close to $(\pi,0)$,
for two different  doping concentrations.
The  $T_{\rm c}=78\, {\rm K}$  sample is slightly overdoped whereas
the  $T_{\rm c}=88\, {\rm K}$ sample is underdoped.}
 \label{fA1}
\end{figure}

In Fig.~\ref{fA1}, we show ARPES spectra close to the momentum ${\bf  k}=(\pi,0)$,
for two different  doping concentrations~\cite{WZX}.
While for the overdoped, $T_{\rm c}=78\, {\rm K}$, sample a rather
sharp peak occurs, which crosses the Fermi energy,  the spectral
density of
the underdoped,  $T_{\rm c}=88\, {\rm K}$, sample  exhibits instead  a very
broad
maximum at approximately $200\, {\rm meV}$. 
Thus,  the entire line-shape
changes character as the doping is reduced.
The other important difference between  the two charge carrier concentrations
is the appearance of the leading edge gap (LEG), i.e. a shift of the
lowest binding energy relative to the chemical potential, for the
underdoped system. This LEG varies between $20$ and $30\, {\rm meV}$
and is therefore hardly visible in Fig.~\ref{fA1} , but is discussed in detail
in Ref.\cite{ZX95,Ding}.

\begin{figure} 
\centerline{\epsfig{file=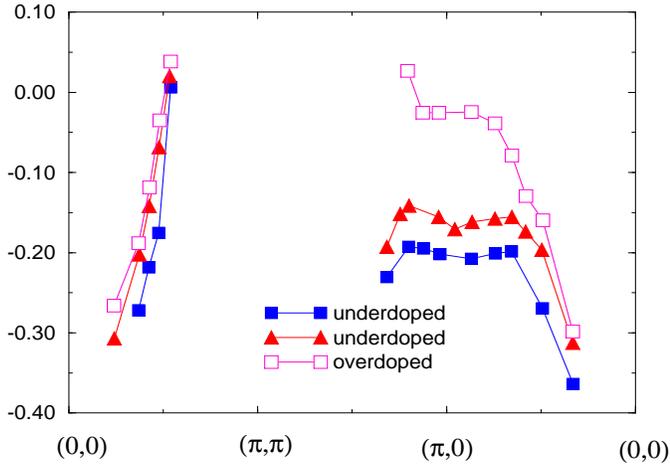,width=10cm, height=8.2cm, scale=1}}
\caption{The position of local maxima of the spectral function along
 the 
 high symmetry lines of the Brillouin zone is shown for an
overdoped and underdoped system (data from Ref. [31]).}
\label{fA2}
\end{figure}

 In addition to  this strong doping dependence, the spectral function of
underdoped systems  is also very anisotropic in momentum space, as
can be seen in Fig.~\ref{fA2}.
Here, the position of local maxima of the spectral function along
certain high symmetry lines of the Brillouin zone is shown for an
overdoped and underdoped system. This  is usually  done because the   maxima
of the spectral density correspond to the position of the quasiparticle
energy. 
However, as we discuss in detail below, 
this  interpretation is not  correct
in underdoped systems for momentum states close to $(\pi,0)$ where     the  
line-shape  changes qualitatively.
Close to $(\pi,0)$, one sees for   overdoped systems,  in agreement
with Fig.~\ref{fA1}, 
a peak at low
binding energy, which crosses the Fermi energy between $(\pi,0)$ and
$(\pi,\pi)$,
whereas the  $200\, {\rm meV}$ high energy feature is the only visible
structure for the underdoped system. It is flat and seems even
repelled from  the Fermi energy 
between $(\pi,0)$ and
$(\pi,\pi)$. 
The situation is different for momentum states along the diagonal,
where a rather sharp peak crosses the Fermi energy between $(0,0)$ and
$(\pi/2,\pi/2)$;  the velocity of the latter  states, seen in the
slope of their dispersion in Fig.\ \ref{fA2}, is independent of  the
doping value, while   no LEG has been observed  for those
quasiparticle states.

\begin{figure}
\centerline{\epsfig{file=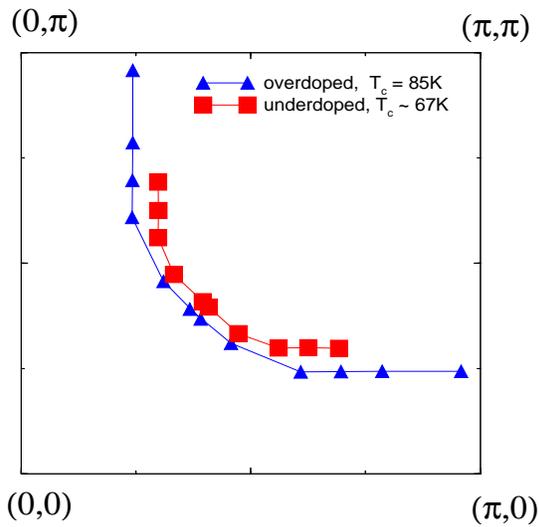,width=9cm, height=7.38cm, scale=1}}
\caption{Fermi surface for an overdoped and underdoped system,
obtained from ${\bf k}$-points where local maxima of the spectral
function cross the Fermi energy (data from Ref. [31]).}
\label{fA3}
\end{figure}

\begin{figure}
\centerline{\epsfig{file=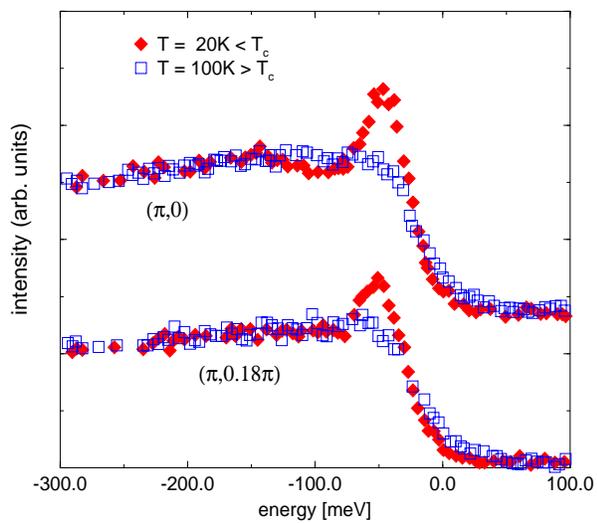,width=9cm, height=8cm, scale=1}}
\caption{Spectral function of a hot quasiparticle above and below the
superconducting transition temperature (data from Ref. [22]).}
\label{fA4}
\end{figure}

In Ref.~\cite{MDL96}, the authors constructed the Fermi surface for the two
doping regimes by determining the ${\bf k}$-points where a maximum of
the spectral function crosses the Fermi energy. Their results are
re-plotted in Fig.~\ref{fA3}.
Consistent with Fig.~\ref{fA2} a large Fermi surface  occurs for the overdoped
material, whereas only a small Fermi surface sector close to the
diagonal could be identified in the underdoped case.
Even though this appears to be in agreement with the formation of a
hole pocket closed around $(\pi/2,\pi/2)$, with reduced intensity on
the other half of the pocket, ARPES data below the superconducting
transition temperature, shown in Fig.~\ref{fA4},  show  that for  momenta
close to $(\pi,0)$, a sharp peak appears at lower binding energy.
This behavior,  for the underdoped
case  is completely consistent with a large Fermi surface  which is
only gapped due to the superconducting state.
The obvious question arises:   how could  a transformation from a small to
a large Fermi surface   occur  on entering the superconducting 
state?

We should also mention that in Ref.~\cite{WZX}, the authors showed that
the two energy scales (the LEG and the high energy feature) behave as
function of 
doping in a   fashion which is quite reminiscent of 
 the two temperature scales,
$T_*$ and $T^{\rm cr}$, shown in Fig.~\ref{fI1}. This leads 
  us immediately to two conjectures:
1. There is a   relationship of the physics of the upper
crossover temperature $T^{\rm cr}$ and the high energy feature,  as well
as between the strong pseudogap temperature $T_*$ and the LEG.
2. As  a  strong pseudogap state is impossible without a weak
pseudogap state at higher temperatures,   the
LEG can only appear after the system has established the high energy
features.

As noted above,  
these fascinating experimental  results represent  a set of very strong
constraints for the microscopic description of underdoped cuprates we
develop below.

\section{the spin fluctuation model}

The nearly antiferromagnetic Fermi liquid (NAFL)
 model~\cite{MP93,MBP93} of the cuprates offers a possible explanation
for the observed weak and strong pseudogap
behavior.
It is based on the spin fluctuation   model, in which   the magnetic interaction between the 
  quasiparticles of the CuO$_2$ planes  is responsible for the 
anomalous normal state properties and the superconducting state with
high $T_c$ and d$_{x^2-y^2}$ pairing state~\cite{MP93,PM95}.
In a recent  letter,   we have    shown how  the weak
pseudogap regime can be understood within this  NAFL-scenario~\cite{SPS97}.

In common with many other approaches,  within the spin fluctuation model 
the planar quasiparticles are assumed to be characterized by a starting spectrum which reflects
their barely itinerant character, and which takes into account both nearest neighbor
and next nearest neighbor
hopping, according to
\begin{equation}
\varepsilon_{\bf k}=-2t(\cos k_x +\cos k_y ) -4t' \cos k_x \cos k_y -\mu \, ,
\label{disp}
\end{equation} 
where $t$, the nearest neighbor hopping term, $ \sim 0.25$ eV,
while the next nearest neighbor hopping term, $t'$,  may vary between 
$t'\approx -0.45t$
for YBa$_2$Cu$_3$O$_{6+\delta}$ and $t' \approx -0.25t $ for La$_{2-x}$Sr$_x$CuO$_4$.

In distinction to many other models, the spin fluctuation model starts from the
ansatz that  
the highly anisotropic
 effective planar quasiparticle interaction mirrors the
 dynamical spin susceptibility~\cite{MMP90}, 
\begin{equation}
  \chi_{\bf q}(\omega)=\frac{ 
 \alpha \xi^2}{1+\xi^2({\bf q}-{\bf Q})^2
- i {\omega\over \omega_{\rm sf}}}\, ,
\label{MMP}
\end{equation}
peaked near ${\bf Q}=(\pi.\pi)$,  via:
 \begin{equation}
V_{\rm eff}^{\rm NAFL}({\bf q},\omega)=g^2 \chi_{\bf q}(\omega)\,  ,
\label{eff_int}
\end{equation}
an ansatz which enables us to construct directly a theory which focuses solely
on the relevant low energy degrees of freedom.
In Eq.~\ref{eff_int}, $g$ is the coupling constant characterizing the 
interaction strength of the  planar  quasiparticles with their own 
collective spin excitations. In this model, changes in quasiparticle behavior
 both reflect and bring about the measured changes in spin dynamics.
The  dynamic susceptibility,  Eq.~\ref{MMP},  was introduced by Millis,
 Monien, and Pines~\cite{MMP90} to explain NMR experiments,
which can be used to determine the correlation length, $\xi$, the 
constant scale factor, $\alpha$, and the  energy scale $\omega_{\rm sf}$, which characterizes
the over-damped nature of the spin excitations.
It follows from the experimental  data that
the static staggered spin susceptibility $\chi_{\bf Q}=\alpha \xi^2$ is large compared to 
the uniform spin susceptibility, $\chi_0$,  and the relaxational mode
energy correspondingly small compared
to the planar quasiparticle band width~\cite{BP95,ZBP96}. 
For optimally doped and underdoped systems one finds that 
over a considerable regime
of temperatures,
\begin{equation}
\omega_{\rm sf} \ll \pi T
\end{equation}
and it is only as $T$ falls below $T_*$  that $\omega_{\rm sf}$ becomes 
comparable to  and  eventually  larger  than $\pi T$.
In detail, between  $T_*$ and $T^{\rm cr}$  one finds: 
$\omega_{\rm sf}/(\pi T) \approx 0.17$
for YBa$_2$Cu$_4$O$_8$    
 and $\omega_{\rm sf}/(\pi T) \approx 0.14$ for
YBa$_2$Cu$_3$O$_{6.63}$ rather independent of $T$~\cite{BP95}.
As a result  of  this comparatively  
low characteristic energy  found in   the weak pseudogap region,  
the spin system, for ${\bf q}\sim {\bf Q}$,
 is thermally excited and behaves quasi-statically~\cite{SPS97};
the   quasiparticles see a
spin system which  acts like a static  deformation potential, 
a  behavior  which is no longer found below $T_*$  where
$\omega_{\rm sf}$ increases rapidly~\cite{BP95} and 
 the lowest energy scale is the temperature itself.

\begin{figure}
\centerline{\epsfig{file=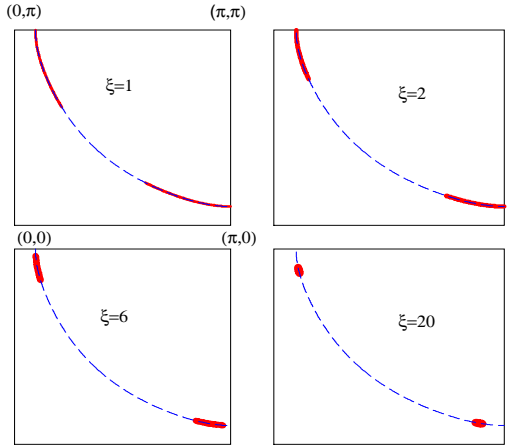,width=9.cm,height=7.38cm, scale=1}}
\caption{A typical  bare Fermi surface
in the first quarter of the BZ, closed around the momentum  point 
$( \pi,\pi)$,     for    different  AF correlation lengths.   The thick
sections   characterize the {\em hot} parts of the Fermi surface. }
\label{fI2}
\end{figure}

 Since the dynamical spin susceptibility $\chi_{\bf q}(\omega)$ peaks  
 at wave vectors
close to $(\pi,\pi)$, two different kinds of 
quasiparticles emerge~\cite{HR,SP96}: {\em hot quasiparticles} with
\begin{equation}
|\varepsilon_{\bf k}-\varepsilon_{{\bf k}+{\bf Q}} |< v/\xi \, ,
\label{hotqp}
\end{equation}
 located 
close to those  momentum points on the Fermi surface which
 can be  connected by ${\bf Q}$,
feel the full effects of the interaction of Eq.\ (\ref{MMP});  {\em cold
quasiparticles} with $|\varepsilon_{\bf k}-\varepsilon_{{\bf k}+{\bf Q}} |> v/\xi$,
 located not far from  the diagonals, 
$\vert k_x\vert=\vert k_y\vert$, feel a ``normal''  interaction.
In Fig.~\ref{fI2}, we show the Fermi surface in the first quarter of
the BZ and indicate the evolution with $\xi$ of  its  hot regions,
  which fulfill Eq.~\ref{hotqp},
by  a  thick line.  
 Note that  even for a correlation length $\xi=1$ a
different behavior along the diagonal and away from it is expected.
For larger  values of $\xi$, the hot regions  become  smaller
while 
their effective interaction  increases.
 Close to $T_{\rm c}$,   typical values for
$\xi$ 
 of underdoped
but superconducting cuprates are $2 < \xi < 8$,  depending on
 doping concentration~\cite{BP95};
 $v$ is the magnitude of a typical Fermi velocity in the corresponding
momentum regions.

The distinct lifetimes of hot and cold quasiparticles  can  be 
obtained  from transport experiments:
 a detailed analysis shows that, due to the almost singular interaction,
the behavior of the hot quasiparticles is highly  anomalous, 
while cold quasiparticles 
 may be characterized as a strongly  coupled Landau Fermi Liquid~\cite{SP96}.
The presence of incommensurate peaks in the
spin fluctuation spectrum~\cite{AMH97,Mook}, and hence in the NAFL interaction,
although difficult to calculate, may be expected to 
amplify the role played by hot quasiparticles in the
determination of system behavior.

   In   the spin fluctuation  model  the anomalous    
 behavior of the cuprates  
is assumed to originate in  
a   strong 
  interaction between fermionic 
spins   ${\bf s}_{\bf q} =\frac{1}{2}\sum_{{\bf k} \sigma \sigma'} 
c^\dagger_{{\bf k}+{\bf q} \sigma} 
{\bf \sigma}_{\sigma \sigma'} c_{{\bf k} \sigma'}$
which brings about intermediate range ($\xi > 1$) antiferromagnetic
spin correlations and  over-damped  spin modes.
   Here,  the     operator
$c^\dagger_{{\bf k}\sigma}$ creates a quasiparticle   
which consists of hybridized copper 3d$_{x^2-y^2}$ and oxygen 2p$_{x(y)}$
states~\cite{ZRsingl}.
The quantity of central physical interest is the dynamical spin susceptibility
\begin{equation}
\chi_{\bf q}(\tau - \tau')= 
\langle T_\tau s^\alpha_{\bf q}(\tau) s^\alpha_{-{\bf q}}(\tau') \rangle \, .
\end{equation}
 which after Fourier transformation in frequency space and analytical
continuation to the real axis is assumed to take the form,  Eq.~\ref{MMP}.
  The intermediate and  low energy degrees of freedom 
are characterized  by  an effective action~\cite{MP93}
\begin{eqnarray}
S&=&-\int_0^\beta d\tau \int_0^\beta d\tau'  \left(
\sum_{{\bf k},\sigma} c^\dagger_{{\bf k}\sigma}(\tau) \, 
G^{-1}_{o {\bf k}}(\tau-\tau') \, c_{{\bf k}\sigma}(\tau') \right. \nonumber \\
& &+\left.  g^2 \frac{2}{3}
 \sum_{{\bf q}} \chi_{\bf q}(\tau -\tau') 
\, {\bf s}_{\bf q}(\tau) \cdot {\bf s}_{-{\bf q}} (\tau') \right)\, ,
\label{act_spinf}
\end{eqnarray}
where $G^{-1}_{o {\bf k}}(\tau-\tau')=
-(\partial_\tau +\varepsilon_{\bf k} )\delta(\tau-\tau')$ 
is the inverse  of the unperturbed  single particle Green's function with the bare dispersion,
 Eq.~\ref{disp}.
In using Eq.~\ref{act_spinf}, we implicitly assume that the effect of
all other high energy degrees of freedom, which are integrated out to
obtain the action $S$, do not affect the Fermi liquid character of the
quasiparticles. In Eq.~\ref{act_spinf},
the effective spin-spin interaction is
assumed to be fully renormalized; thus it reflects the changes in
quasiparticle behavior it brings about, and can be taken from fits to
NMR and  INS experiments.
We will also assume that the spin degrees of freedom are completely
isotropic and that  all three components of the spin vector are equally 
active. In the case of  a  intermediate  correlation 
lengths  $1 \leq  \xi  \leq 8 $, 
  this  is  the appropriate description
of the spin  degrees of freedom. Only for  much larger $\xi \propto
\exp(const./T)$,  does one enter  the regime in which   even without long
 range order only two transverse spin  degrees of freedom are
 active~\cite{CM97}. The physics of the crossover,  
driven by a collective-mode collective-mode interaction,
between these two regimes,  is beyond the
scope of this paper.
 
The quantities of   primary interest to us are the single particle Green's function
$G_{{\bf k},\sigma} (\tau - \tau')=
-\langle T c_{{\bf k} \sigma}(\tau) c^\dagger_{{\bf k} \sigma}(\tau') \rangle$
which provides information about the quasiparticle spectral density  determined
in angular resolved photo-emission experiments,  the  dynamical
spin susceptibility itself, and the corresponding charge response
functions.
As noted above, in calculating these quantities for intermediate
correlation lengths 
  the interaction  between the 
collective spin modes is irrelevant.
In appendix~\ref{App_GF} we   show  that  under these circumstances
the Green's function 
\begin{equation}
G_{{\bf k} \sigma}(\tau-\tau')=   \left\langle
 \hat{G}_{{\bf k},{\bf k} \sigma \sigma }(\tau,\tau' |{\bf S} ) \right\rangle_o  
\label{GFapprox_main}
\end{equation}
can be  expressed  as  a Gaussian average $\langle \cdots \rangle_o $ of  electron propagators
 with a  time dependent magnetic potential $\frac{g}{\sqrt{3} }{\bf S}_{\bf q}(\tau)$, 
with respect to collective bosonic spin 1 variables ${\bf S}_{\bf q}(\tau)$.
The corresponding model is often referred to as the spin fermion
model.
We give in appendix~\ref{App_GF} 
 the   diagrammatic rules of this   problem,
which will be essential for the   solution of the 
spin fermion model in the quasistatic limit.
In the next  two sections, we derive   new   expressions for   the
single particle Green's function, and the spin-fermion  and charge-fermion 
vertex functions of the quasistatic two 
dimensional spin fermion model,  
valid for intermediate values of the spin fermion coupling,
by extending an earlier study
by Sadovskii~\cite{Sad79} for   one dimensional
 charge density wave systems.

\section{Theory of the quasistatic limit}
We begin this section by first motivating  the quasistatic limit and
discussing its   physical
consequences by  investigating   the second
order diagram with respect to the coupling constant $g$.
 We then  present a solution of the spin fermion model
which is not restricted to the  weak coupling regime and provides new
insight
into the  intermediate coupling behavior relevant for underdoped
cuprates.

 \subsection{The second order diagram and the static limit}
In second  order perturbation theory, the quasiparticle self energy is
given by:
\begin{equation}
\Sigma_{\bf k}(i\omega_n)=g^2 \frac{1}{\beta} \sum_{{\bf q},m}
\chi_{\bf q}(i\nu_m) \, \frac{1}{i\omega_n+i\nu_m -\varepsilon_{{\bf k} + {\bf q}}}\, .
\label{self2otot}
\end{equation}
If, for a given temperature $T$,
 the characteristic frequency of the spin excitations $\omega_{\rm sf}$ is small
compared to the intrinsic thermal broadening of the electronic states,
the energy transfer $i\nu_m$ of this state due to an inelastic scattering process
is negligible. Furthermore, in the limit $\pi T \gg \omega_{\rm sf}$, 
$\chi_{\bf q}(i\nu_m)$ is dominated by the Matsubara frequency $\nu_m=0$, 
 so Eq.~\ref{self2otot} takes the form:
 \begin{equation}
\Sigma_{\bf k}(i\omega_n)=
\tilde{g}^2   \sum_{{\bf q}}
  S(q)  \frac{1}{i\omega_n-\varepsilon_{{\bf k} +{\bf Q}+ {\bf q}}} \, ,
\label{self2o}
\end{equation}
with $\tilde{g}^2=g^2 \alpha T$ and  
 \begin{equation}
S(q)=\frac{1}{\/\xi^{-2} +  q^2 }  \, .
\label{mom}
\end{equation}
Physically, this use of a 
static approximation  for the  spin degrees of freedom
reflects the fact  that since  the  frequency variation of $\chi_{\bf q}(\omega)$ takes place
on the scale $\omega_{\rm sf}$, once $\pi T \gg \omega_{\rm sf}$, all relevant
collective spin degrees of freedom are thermally excited and the
phase space restrictions for scattering phenomena due to the quantum
mechanical
nature of the spins are irrelevant. 
It follows that 
     we can then neglect the
$\omega$-variation of
$\chi_{\bf q}(\omega)$.

For the system we study, experiment shows that
the dominant momentum transfer  ${\bf q}$ of the spin fluctuations is 
close to the antiferromagnetic wave vector ${\bf Q}=(\pi,\pi)$, 
so that 
  we can  expand
the energy dispersion as
\begin{equation}
\varepsilon_{{\bf k} + {\bf Q}+ {\bf q}} \approx 
\varepsilon_{{\bf k} + {\bf Q}} + {\bf v}_{{\bf k} + {\bf Q}} \cdot {\bf q}
 \label{disp_exp}
\end{equation}
with velocity $v^\alpha_{{\bf k} + {\bf Q}}=\partial \varepsilon_{{\bf k} + {\bf Q}}/
\partial k_\alpha$.
Note that in distinction to  a one 
  dimensional problem,   the linearization of the electron spectrum   in  two dimensions is not
  straightforward. In Eq.~\ref{disp_exp}, we have 
linearized with
  respect to the {\em transferred} momentum  ${\bf q} \approx {\bf
    Q}=(\pi,\pi)$, an approximation which is   justified  provided   
${\bf q}$  deviates only slightly from the antiferromagnetic
wave vector ${\bf Q}$, i.e. for systems with a sufficiently large 
antiferromagnetic correlation length $\xi$. Therefore, technically
$\xi^{-1}$ is considered to be a small quantity and all related
momentum integrals are evaluated  accordingly.. On comparing this
approximate
treatment with a complete numerical evaluation, we find that it can be
applied  once  $\xi >1$.
 At ${\bf k}=(\pi,0)$, the velocity $v^\alpha_{{\bf k}}$ vanishes
and one must  take   higher order terms  in ${\bf q}-{\bf Q}$ into account.
We   assume  that the physics of this van Hove singularity is irrelevant
(due to three dimensional effects and the presence of  possible additional scattering mechanisms)
and introduce a lower velocity cut off 
$v_c\approx \langle v_{\bf   k}\rangle_{\rm FS}$.
The remaining momentum integration can then  easily be carried out. It
follows,  after analytical continuation $i\omega_n \rightarrow
  \omega+i0^+$, that:
\begin{eqnarray}
 \Sigma_{\bf k}(\omega)&=&\frac{-\Delta^2}
{\sqrt{(v_{{\bf k} + {\bf Q}}/\xi)^2+(\omega-\varepsilon_{{\bf k}+{\bf Q}})^2}}
\left(i\frac{\pi}{2} \right. \label{self_sec_cal} \\
& &\left. - {\rm arctanh}\left(\frac{\omega-\varepsilon_{{\bf k}+{\bf Q}}}
{\sqrt{ (v_{{\bf k} + {\bf Q}}/\xi)^2+(\omega-\varepsilon_{{\bf k}+{\bf Q}})^2} }\right) 
\right)\, ,  \nonumber
 \end{eqnarray}
where $\Delta^2= g^2 \alpha T \log(1+(\xi\Lambda)^2)$ and  
$\Lambda \approx \pi$ is  the  upper cut off of the momentum summation.
Since we are technically at high temperatures, our results  
depend on this cut off, which is undesirable. 
We avoid 
  this
problem by expressing any    cut-off dependence of the theory in terms
of measurable quantities: Thus on using the local moment sum rule 
$\langle {\bf S}_i^2 \rangle =3 T \sum_{m,{\bf q} } \chi_{\bf q}(i
\nu_m)$, 
we find that
$\Delta$ can also be expressed
as
\begin{equation}
\Delta^2= g^2 \langle {\bf S}_i ^2 \rangle /3 \, .
\end{equation}
We therefore  can  use  this 
expression for $\Delta$ and  determine $\langle {\bf S}_i ^2 \rangle$
from the experimentally determined susceptibility  $\chi_{\bf
  q}(i\nu_m)$ of Eq.~\ref{MMP}.
 This guarantees    a reasonable estimate for the
   total spectral weight of the spin excitation spectrum
for the spin fluctuation induced scattering processes.
 
Consider a given ${\bf k}$-point on the Fermi surface
($\varepsilon_{\bf k}=0$). If the Fermi surface is such that the
momentum transfer by ${\bf Q}$ takes you to another Fermi surface
point, i.e. $\varepsilon_{{\bf k}+{\bf Q}}=0$, it follows from 
Eq.~\ref{self_sec_cal}  that for this momentum state,
 a so called  {\em hot spot},
$\varepsilon_{{\bf k}}=\varepsilon_{{\bf k}+{\bf Q}} =0$,  
the  real part of the self energy decreases like ${\rm log}(\omega)/\omega$ 
if $\omega > v/\xi$, close to the $1/\omega$
 behavior which 
  is a signature of  precursor effects of  an spin density
  wave~\cite{TT1_a}.
More generally, 
  anomalous scattering processes  will continue to  modify the single particle spectrum
 dramatically  for those  momentum states for which
\begin{equation}
|\varepsilon_{{\bf k}}-\varepsilon_{{\bf k}+{\bf Q}}| < v/\xi\, .
\label{hotcrit}
\end{equation}
This entire region of the BZ behaves in  qualitatively different
fashion  from the rest of the system; it 
  corresponds to the definition of  {\em hot} quasiparticles discussed recently by
Stojkovi\'c and Pines~\cite{SP96}.

We call attention to the fact that only for the hot quasiparticles can
we justify neglecting the higher Matsubara frequencies.
For cold
quasiparticles with $ |\varepsilon_{{\bf k}}-\varepsilon_{{\bf
    k}+{\bf Q}}| > v/\xi$ the characteristic energy scale of the spin
fluctuations
is no longer  $\omega_{\rm sf}$ but  turns out to be  $\sim
\omega_{\rm sf} \xi^2$~\cite{SP96},
a quantity 
which is not, in general,    small compared to $\pi T$. 
As a result,  our approach, while properly accounting 
for the anomalously large scattering rate and related new physics of
the hot quasiparticles,  will  tend  to
overestimate
the scattering rate for cold quasiparticles. 
Put another way, differences in behavior between hot and cold
quasiparticles will be underestimated in our theory.

In order to  make explicit the role played by the  presence 
  of SDW precursors in the  
 quasistatic  regime,  we
  evaluate the above  momentum integrals  within  the  approximation
\begin{equation}
S(q)  \approx \frac{\/\xi^{-1}}{\/\xi^{-2} +  q_\parallel^2} \, 
\frac{\/\xi^{-1}}{\/\xi^{-2} + q_\perp^2} \, ,
\label{momapp}
\end{equation}
where $q_{\parallel (\perp)}$ is the projection of ${\bf q}$  parallel (perpendicular) to 
 the velocity 
$ {\bf v}_{{\bf k} + {\bf Q}} $.
 We then obtain:
\begin{eqnarray}
\Sigma_{\bf k}(\omega)  =
\frac{\Delta^2}{\omega -\varepsilon_{{\bf k}+{\bf Q}}+i v_{{\bf k}+{\bf Q}}/\xi }\, ,
\label{seco}
\end{eqnarray}
an expression which, apart from a logarithm, has 
  the same anomalous behavior as Eq.~\ref{self_sec_cal}.
In the limit $\xi \rightarrow \infty$
$\Delta$ is the spin density wave gap and 
 the poles of the resulting   Green's function 
are the two branches of  the mean field SDW  state discussed by Kampf
and Schrieffer~\cite{TT1_a}.

For the investigation of higher order diagrams in the next paragraph, 
it  will be    helpful to introduce the 
following representation of the second order self energy:
\begin{equation}
\Sigma_{\bf k}(\omega) =  -i \tilde{g}^2  \int_0^\infty dt\, e^{i(\omega -\varepsilon_{{\bf k}+{\bf Q}})t}
 \psi_{{\bf k}+{\bf Q}}(t)\, 
\end{equation}
where
\begin{equation}
\psi_{{\bf k}+{\bf Q}}(t)=\sum_{\bf q} 
\frac{1}{\/\xi^{-2} +  q^2 }  e^{-i {\bf v}_{{\bf k}+{\bf Q}}  \cdot {\bf q}  t}\, .
\label{psidef}
\end{equation}
Evaluation of the momentum summation yields for $\Lambda \rightarrow \infty$:
 \begin{equation}
\psi_{{\bf k}+{\bf Q}}(t)=2 \pi K_0(t v_{{\bf k}+{\bf Q}}/\xi)\, ,
\label{psi1}
\end{equation}
where $K_0$ is the modified Bessel function.
Using the approximation of Eq.~\ref{momapp}, this simplifies to
\begin{equation}
\psi_{{\bf k}+{\bf Q}}(t)\approx e^{-t v_{{\bf k}+{\bf Q}}/\xi} \, .
\label{psi2}
\end{equation}

The  tendency   towards   SDW   behavior in the quasi-static  regime
so far relies
on the applicability of the second order  perturbation theory:
 visible effects can only occur once the correlation length exceeds the
electronic length scale $\xi_o =v/\Delta \approx  2v/g$. In a weak coupling treatment,
the  above discussion is applicable 
only for large correlation length:    one therefore
 has to go
beyond second order  perturbation theory to be certain     whether or not
SDW precursors are relevant for  cuprates with intermediate
correlation length.  This is possible only if $\xi_o$  is only a few
lattice constants; it    implies that we have to investigate an
intermediate coupling regime.
Therefore, we  present in the next paragraph a procedure which enables
us to sum the entire perturbation series.

\subsection{  Diagram summation in the quasistatic limit }
To 
 evaluate all higher order self energy diagrams    within the
 quasistatic limit,  we
 first derive a compact expression for an arbitrary diagram 
and then,
 as 
  a second step, sum {\em all}
diagrams of the perturbation series to obtain  the self energy and 
single particle Green's function.
This summation is made  possible by the fact 
that many diagrams with
rather different topology  are,  apart from a factor which describes
 multiplicity and sign, identical.

\begin{figure}
\centerline{\psfig{file=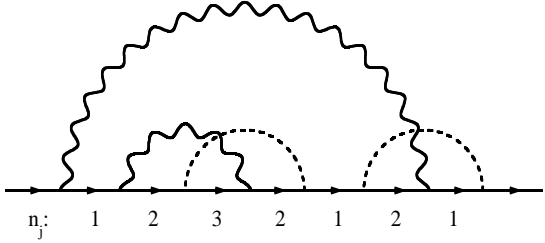,width=8cm, height=4.5cm, scale=2}}
\caption{Illustration of the sequence $\{n_j \}$ for a self energy
  diagram of order $g^8$. $n_j$ is the number of spin fluctuation
  lines above the $j$-th fermionic Green's function. }
\label{fT1}
\end{figure}

As  first shown
by Elyutin  in the context of optical  response in a
random radiation field~\cite{Ely76}, diagrams can be characterized by
the sequence of integer 
numbers $\{n_j\}$, where
 $n_j$ is  the number of interaction lines above the $j$-th Green's
 function;  for an example,   see
  Fig.~\ref{fT1}.
In the following we  prove 
that in the quasistatic regime, 
diagrams with the same sequence $\{n_j\}$ are proportional to each other.
 The proportionality factor        will be  determined  below.

 An arbitrary diagram of order $2N$ can, up to a constant,
      be expressed as:
\begin{eqnarray}
 \Sigma^{(2N)}({\bf k},\omega) &= & \tilde{g}^{2N} \sum_{{\bf q}_1 \cdots {\bf q}_N} 
S( \tilde{q}_1)  \cdots S( \tilde{q}_N) 
\nonumber \\  
& & 
\prod_{j=1}^{2N-1} G_{o, {\bf k}+ \sum_{\alpha=1}^N 
R_{j,\alpha} {\bf q}_\alpha}(\omega)\, ,
\label{self_N}
\end{eqnarray}
where $\tilde{\bf q}_\alpha={\bf q}_\alpha-{\bf Q}$  and the $\left( (2N-1) \times N \right)$ 
matrix $R_{j, \alpha}$ determines whether   ${\bf q}_\alpha$
($\alpha=1 \ldots N$) occurs as a momentum 
 transfer in the $j$-th Greens function ($j=1 \ldots 2N-1$)
 of  the diagram, i.e. $ R_{j, \alpha}=1$ or $0$.
In general, each diagram is  fully characterized by $R_{j, \alpha}$.
It is important to notice that  $n_j$  is given by the expression, 
\begin{equation}
n_j=\sum_{\alpha=1}^N R_{j,\alpha} \, .
\label{sequence}
\end{equation}

Since each of the momenta ${\bf q}_\alpha$ of Eq.~\ref{self_N} is separately constrained 
to  lie in a 
 region close to ${\bf q}_\alpha \approx {\bf Q}$, we  can 
expand:
\begin{eqnarray}
\varepsilon_{{\bf k}+ \sum_{\alpha=1}^N 
R_{j,\alpha} {\bf q}_\alpha }&\approx &\varepsilon_{{\bf k}+j{\bf Q}} \nonumber \\
& &+{\bf v}_{{\bf k}+j{\bf Q}}
\sum_{\alpha=1}^N  R_{j,\alpha} \left( {\bf q}_\alpha-{\bf Q}
\right)\, ,
\label{lin_disp_g}
\end{eqnarray}
where we have used the fact that  $n_j$ is even (odd) if  $j$ is even
(odd)  
since at each vertex $n_j$ 
changes by $\pm 1$ and $n_1=1$.
Shifting all momenta   ${\bf q}_\alpha-{\bf Q} \rightarrow {\bf q}_\alpha$ and   
introducing, as we have done for the 
second order diagram, $2N-1$ auxiliary time variables $t_j$,  it
follows that:
\begin{eqnarray}
& &\Sigma^{(2N)}({\bf k},\omega) = (-i)^{2N-1}\, \tilde{g}^{2 N }
\sum_{{\bf q}_1 \cdots {\bf q}_N}  S(q_1) \cdots S(q_N)  \nonumber \\
\lefteqn{\prod_{j=1}^{2N-1} \int_0^\infty\, dt_j\,
e^{i t_j \left(\omega -\varepsilon_{{\bf k}+j{\bf Q}} 
   -  {\bf v}_{{\bf k},j}   \sum_{\alpha=1}^N R_{j,\alpha} 
 {\bf q}_{ \alpha}   \right)} \, ,}  
\end{eqnarray}
 with   ${\bf v}_{{\bf k},j} ={\bf  v}_{{\bf k}+j{\bf Q}}$.
In this 
 proper time representation of the self energy, the different 
momentum integrals decouple; on 
  using
   Eq.~\ref{psidef} it follows that 
\begin{eqnarray}
\Sigma^{(2N)}({\bf k},\omega) &=& (-i)^{2N-1}\,\tilde{g}^{2N}\prod_{j=1}^{2N-1}\int_0^\infty\, dt_j\,
e^{i(\omega -\varepsilon_{{\bf k}+j{\bf Q}})t_j}\nonumber \\ & &
\times \prod_{\alpha=1}^{N} \psi_{{\bf k}+j{\bf Q}} ( R_{j,\alpha} t_j ) \, .
\label{aaa}
 \end{eqnarray}
In the last step we   used the fact 
 that    the momentum transfer is sufficiently 
 close to ${\bf Q}$ that 
 we can  neglect   contributions of order $({\bf q}-{\bf Q})^4$ in
 $S({\bf q})$,
i.e. we have 
 assumed that $\psi_{\bf k}(x+y) \propto \psi_{\bf k}(x) \psi_{\bf
  k}(y) $ as  follows from the use of $S(q)$ in Eq.~\ref{momapp}.
This is consistent with the restriction to momentum transfers close to
${\bf q} \sim {\bf Q}$  which motivated 
the linearization of the electron
spectrum in Eq.~\ref{lin_disp_g}. 
Eq.~\ref{aaa} is only valid for hot spots  where 
 the velocities
 ${\bf v}_{\bf k}$ and ${\bf v}_{{\bf k}+{\bf Q}}$ are
 almost perpendicular
to each other.  
For cold quasiparticles this condition 
 is not fulfilled,  so that 
 our theory   can only  give
a qualitative account for their
 scattering processes.
It has been recently pointed out by Tchernyshyov~\cite{oleg} that Eq.~\ref{aaa}
is in fact not fulfilled in the original one dimensional
solution of Ref.~\cite{Sad79}. Therefore, 
the ideas developed in  Ref.~\cite{Sad79} seem to be much more appropriate 
for our two dimensional case.

Since $R_{j,\alpha}$ is either $0$ or $1$, it follows immediately from Eq.~\ref{aaa}:
\begin{eqnarray}
\prod_{\alpha=1}^{N} \psi_{\bf k}( R_{j,\alpha}t_j ) =\psi_{\bf k}(0)^{N-n_j}
  \psi_{\bf k}(t_j)^{n_j}\, ,
 \end{eqnarray}
with $n_j$  given by  Eq.~\ref{sequence}.
Inserting this result and collecting all the prefactors, if follows~\cite{note1}:
\begin{eqnarray}
\Sigma^{(2N)}({\bf k},\omega) &=& (-i)^{2N-1}\,\Delta^{2N}\prod_{j=1}^{2N-1}\int_0^\infty\, dt_j\,
e^{i(\omega -\varepsilon_{{\bf k}+j{\bf Q}})t_j}\nonumber \\ & &
\times  \psi_{{\bf k}+j{\bf Q}}  (t_j )^{n_j}\, .
 \end{eqnarray}
which proves that a given diagram of order $2N$ is fully determined 
by the sequence $\{n_j\}$  as well as provides an explicit expression for these diagrams.

On making use of 
 the simplified evaluation of the momentum integrals, 
   Eq.~\ref{momapp}, it follows
 with the help of  Eq.~\ref{psi1}, that 
 \begin{equation}
\Sigma^{(2N)}({\bf k},\omega) =\Delta^{2N}\prod_{j=1}^{2N-1} 
\frac{1}{\omega -\varepsilon_{{\bf k}+j{\bf Q}}
+i \,n_j v_{{\bf k} ,j} /\xi}\, ,
\label{sigarb}
\end{equation}
a result which is useful in determining the multiplicity of a given diagram.
  
For the actual evaluation of all diagrams, it is essential that for
each 
sequence $\{n_j\}$,  there is a unique mapping to 
a diagram without crossing interaction lines, since for each $\{n_j\}$
there exists one and only one diagram without crossing interaction lines.
Note, unique is meant in the sense of 
 the topology of a
diagram, not whether it contains longitudinal or transverse spin
fluctuations;  for details see Appendix B.
 This is illustrated for two cases in Fig.~\ref{fT2}, where we show two
self energy diagrams of order $g^4$  and $g^8$, which are,
 within the quasistatic approximation identical  apart from a
 proportionality factor.
>From these considerations, it follows that it  suffices to sum up only
the 
non-crossing diagrams taking into account the identical 
crossing diagrams by their   proper multiplicity factors.
The remaining problem is to determine, for a given order in the coupling constant,
 how many identical diagrams exist. In the case of charged and
 uncharged bosons, 
this problem has been
 solved  by Sadovskii~\cite{Sad79}.
The generalization (see Appendix~\ref{App_GF}) to the case of 
  spin fluctuations is not  
straightforward,   because of the additional $(-1)^C$ factor  of
 crossed spin conserving and spin flip lines.

In  appendix~\ref{App_mult}   we  derive the multiplicity of a given
class of diagrams, i.e.   the number of identical diagrams of a given order 
of the perturbation series, by solving the problem
in the special case $\xi \rightarrow \infty$ and using the fact  that  the  combinatorics of the diagrams
does not  depend on this limit.
Having determined these multiplicities, it is possible to sum the entire perturbation series analytically.
We find the following  recursion relation for  the  Green's function
 $G_{\bf k}(\omega) \equiv G^{(j=0)}_{\bf k}(\omega)$:
\begin{equation}
{G^{(j)}_{\bf k}(\omega)}^{-1}=  g^{(j)}_{\bf k}(\omega)^{-1}-
\kappa_{j+1} \Delta^2 G^{(j+1)}_{\bf k}(\omega).
\label{sadself}
\end{equation}
with  $\kappa_j=(j+2)/3$ 
if $j$ is odd and $\kappa_j=j/3$ if $j$ is even
and
\begin{equation}
g^{(j)}_{\bf k}(\omega) =-i \int_0^\infty dt \, 
e^{i(\omega-\varepsilon_{{\bf k}+j{\bf Q}})}\,  \psi_{{\bf k}+j{\bf Q}}(t)^j\, .
\end{equation}
Eq. \ref{sadself} is one of the central results of our theory.  
 This  recursion relation, closed by $G^{(L)}_{\bf k}(\omega)=
g^{(L)}_{\bf k}(\omega)$ for some large value of $L$,
enables us 
to calculate the single particle spectral function $A({\bf k},\omega)$ to arbitrary order $2L$  in
the coupling constant $g$ (we use $L \sim 10^2-10^3$; Eq.~\ref{sadself}
converges for  $L \sim 10^2$).

\subsection{  Spin susceptibility and  vertex function }
Within the  quasistatic limit of the effective low energy quasiparticle interaction,
  we can obtain  an exact
expression for the irreducible part of the dynamical spin susceptibility
and the electron spin fluctuation vertex.
Note that   we are not able to calculate the total susceptibility. 
Since we are
assuming that the interaction line is given by the fully renormalized
spin susceptibility, a direct approach  would  lead to an over counting
of diagrams. Therefore, we   only calculate 
the irreducible part  $\tilde{\chi}^{\alpha \beta}_{\bf q}(i\omega_n)$ 
of the total susceptibility  $\chi^{\alpha \beta}_{\bf q}(i\omega_n)$.
 The latter   can be expressed as  
\begin{equation}
\chi^{\alpha \beta}_{\bf q}(i\omega_n)=\frac{\tilde{\chi}^{\alpha \beta}_{\bf q}(i\omega_n)}
{1-f_{\bf q} \tilde{\chi}^{\alpha \beta}_{\bf q}(i\omega_n)} \,  ,
\end{equation}
where the restoring force, $f_{\bf q} $, is determined by the
renormalization of the 
spin exchange fermion-fermion
interaction through high energy excitations in all other channels.
 $f_{\bf q} $ is   then  related in a non-trivial fashion to the
 underlying 
microscopic Hamiltonian of the system and has to be considered as an
additional input quantity. 

By following a procedure analogous to that in determining 
the  Green's function in 
Eq.~\ref{GFapprox_main}, one can 
show that   the irreducible part of the dynamical spin susceptibility
is given by
\begin{equation}
\tilde{ \chi}^{\alpha \beta}_{\bf q}(i\omega_n)=  \left\langle\, 
\Pi^{\alpha \beta}_{\bf q}(i\omega_n |{\bf S})\,  \right\rangle_o\, ,
\label{irrsus0}
 \end{equation}
where $\Pi^{\alpha \beta}_{\bf q}(i\omega_n |{\bf S}) $ 
is the    irreducible particle hole propagator  
for a given spin field configuration.
\begin{eqnarray}
\Pi^{\alpha \beta}_{\bf q}(i\omega_n |{\bf S})&=& -\frac{1}{4}
\sum_{{\bf k}{\bf k'}m m'} {\rm Tr }\left\{ \, {\bf \sigma}^\alpha 
\hat{G}_{{\bf k}+{\bf q},{\bf k'}}(i\Omega_{m,n}, i\omega_{m'}|{\bf S} ) \right.
\nonumber \\ 
& & \left. \times {\bf \sigma}^\beta
\hat{G}_{{\bf k}'-{\bf q},{\bf k}}(i\Omega_{m',n}, 
i\omega_{m}|{\bf S} )\right\}\, .
\label{pobub}
\end{eqnarray}
Here  ${\rm Tr }\, \ldots$ refers only to the trace in spin space and 
$\Omega_{m,n}=\omega_m+\omega_n$
This result is obtained by neglecting all reducible contributions
  in taking the  functional derivative  with respect to an external 
time dependent magnetic field coupled to the  electron spins ${\bf s}_i(\tau)$.
 
The diagrammatic rules described in  appendix~\ref{App_GF} 
for the single particle Green's function $G_{\bf k}(\omega)$
can  be extended  in straightforward fashion to the spin susceptibility, 
which   can be expressed in terms of $G_{\bf k}(\omega)$ and the electron spin
 fluctuation vertex function:
\begin{eqnarray}
\tilde{\chi}_{\bf q}(i\nu_m)&=&\frac{1}{\beta} \sum_{{\bf k},n} G_{\bf k}(i \omega_n)
G_{{\bf k}+{\bf q}}(i \omega_n+i\nu_m )\nonumber \\ 
& &  \Gamma^{\rm s}_{{\bf k},{\bf k}+{\bf q}}(i \omega_n,i \omega_n+i\nu_m )\, .
\label{susvertex}
\end{eqnarray}
Thus,  a  knowledge of the 
vertex function gives immediate information about the irreducible part of the 
dynamical spin susceptibility. A similar relation exists for the
corresponding charge susceptibility.
In this paragraph we outline the exact determination of 
$\Gamma^{\rm s (c)}_{{\bf k},{\bf k}+{\bf q}}(\omega+i0^+,\omega+\nu+i0^+)$
 obtained after analytical continuation
to the real axis. For the  determination of the susceptibility on the real axis we will also need
the  analytical continuation
 $\Gamma^{\rm s (c)}_{{\bf k},{\bf k}+{\bf q}}(\omega-i0^+,\omega+\nu+i0^+)$
which has to be determined independently but 
can be obtained in a similar way.

 As was the case for    the electronic Green's function,  the vertex
 function  is  obtained in two steps: first,  based on purely diagrammatic arguments we obtain
a general expression for the vertex function in terms of the 
previously determined Green's function and some combinatorial 
prefactors which  take  the  proper multiplicity of the diagrams
into account; second,    these prefactors are determined in 
the limit $\xi \rightarrow \infty$. This is possible because
the combinatorics of the diagrams does not depend on the actual value
of $\xi$.
Finally, we obtain    a closed expression valid  for all values of $\xi$.

In the case of the Green's function 
each diagram was  proportional  to a rainbow
diagram. The corresponding conclusion for the vertex function is that 
each vertex diagram is identical to  a  diagram of  the ladder
approximation and the entire perturbation series can be obtained by 
summing    the ladder series   with appropriate weighting factors.
The proof of this statement is almost identical to the corresponding proof for the 
Green's function.  
 Because  an arbitrary diagram can be related  to  a ladder  diagram, it follows that 
the vertex
 $\Gamma^{\rm s (c)}_{{\bf k},{\bf k}+{\bf q}}(\omega+i0^+,\omega+\nu+i0^+)
\equiv \Gamma^{(0), {\rm s (c)}}_{{\bf k},{\bf k}+{\bf q}}
(\omega+i0^+,\omega+\nu+i0^+)$
can be generated by two renormalized Green's functions and an 
effective vertex 
$\Gamma^{(1), {\rm s (c)}}_{{\bf k},{\bf k}+{\bf
    q}}(\omega+i0^+,\omega+\nu+i0^+)$ 
which includes all  those  processes
not taken into account by one spin fluctuation propagator crossing the external
 bosonic line. For  the spin vertex, we find the recursion relation,
\begin{eqnarray}
 & &\Gamma^{(0), {\rm s}}_{{\bf k},{\bf k}+{\bf q}}
(\omega+i0^+,\omega+\nu+i0^+) =1-r_1 \Delta^2 
G^{(1)}_{\bf k}(\omega)  
\nonumber \\
\lefteqn{\times G^{(1)}_{{\bf k}+{\bf q}}(\omega+\nu)
 \Gamma^{(1) {\rm s}}_{{\bf k},{\bf k}+{\bf
     q}}(\omega+i0^+,\omega+\nu+i0^+)\, .}
\label{V_11}
\end{eqnarray}
Here, the Green's function $G^{(1)}_{\bf k}(\omega)$, Eq.~\ref{sadself},
takes into account that for the diagram under consideration one 
has at least one interaction line 
above each fermionic propagator~\cite{Sad79}.
A comparison with  perturbation theory shows that 
the multiplicity factor which enters  Eq.~\ref{V_11} is given by 
 $r_1=\frac{1}{3}$.  
The minus sign in Eq.~\ref{V_11} results 
  from the diagrammatic rules of appendix~\ref{App_GF}.
 Since the higher order vertex function can be determined 
in exactly the same way as Eq.~\ref{V_11}, one obtains the  recursion
relation
\begin{eqnarray}
 & &\Gamma^{(l), {\rm s}}_{{\bf k},{\bf k}+{\bf q}}
(\omega+i0^+,\omega+\nu+i0^+) =1-r_{l+1} \Delta^2 
G^{(l+1)}_{\bf k}(\omega)  
\nonumber \\
\lefteqn{\times G^{(l+1)}_{{\bf k}+{\bf q}}(\omega+\nu) 
\Gamma^{(l+1), {\rm s}}_{{\bf k},{\bf k}+{\bf q}}(\omega+i0^+,\omega+\nu+i0^+)\, ,}
\label{vertex}
\end{eqnarray}
which can be  
evaluated using the Green's functions from Eq.~\ref{sadself} and a 
starting
value $ \Gamma^{(L)}=1$.
In Fig.~\ref{fT3}  the diagrammatic motivation for   this  recursion
relation is given: there one sees
  that the problem is  similar to the summation of the
ladder series  for  the vertex, with the difference that all
non-ladder diagrams are taken into account
  by the corresponding weighting factors $r_{l+1}$
Once these  prefactors  are known,   the
vertex function can be determined 
up to arbitrary order of the coupling constant.
The $r_l$ are  defined diagrammatically  by the fact that $3^L \prod_{l=1}^L r_l$
is the number of skeleton diagrams of order $g^{2L}$ which 
contribute to the vertex  function. 
[Note,  that  non-skeleton diagrams are diagrams with interaction
lines which   
only  renormalize   the Green's functions.]

The combinatorial determination of the $r_l$ is  somewhat  cumbersome.
 We proceed by using the general expression, Eq.\ref{vertex},
to calculate the irreducible susceptibility   given in
Eq.~\ref{susvertex},
while determining 
  the  irreducible susceptibility independently 
in the limit $\xi \rightarrow \infty$ analytically by evaluating the path integral
of Eq.~\ref{irrsus0}. On 
comparing these two results for  $\xi \rightarrow \infty$ order by
order in the
 coupling constant we are able to determine 
  the prefactors $r_l$.
On carrying out this calculation for arbitrary momentum ${\bf q}$, we
find 
$r_l=l$ if $l$ is even and $r_l=(l+2)/9$ if $l$ is odd.
 
This completes the specification of  the vertex
function,  Eq.\ref{vertex}, of the spin fermion model and enables us
to calculate
both  the 
irreducible spin susceptibility, $\tilde{\chi}_{\bf q}(\omega)$,   and
the effective 
spin fluctuation induced pairing interaction.

An identical procedure can be performed for the charge vertex
 $\Gamma^{ {\rm c}}_{{\bf k},{\bf k}+{\bf q}}
(\omega+i0^+,\omega+\nu+i0^+) $.
One obtains in place  of Eq.\ref{vertex} the result:
\begin{eqnarray}
 & &\Gamma^{(l), {\rm c}}_{{\bf k},{\bf k}+{\bf q}}
(\omega+i0^+,\omega+\nu+i0^+) =1+\kappa_{l+1} \Delta^2 
G^{(l+1)}_{\bf k}(\omega)  
\nonumber \\
\lefteqn{\times G^{(l+1)}_{{\bf k}+{\bf q}}(\omega+\nu) 
\Gamma^{(l+1), {\rm c}}_{{\bf k},{\bf k}+{\bf q}}(\omega+i0^+,\omega+\nu+i0^+)\, ,}
\label{vertex_C}
\end{eqnarray}
since for the charge vertex  $r_l$  is replaced by  $-\kappa_l $, with $\kappa_l $
as
given in the line below Eq.~\ref{sadself}.

\section{quasiparticle properties: theory compared with experiment}

We consider first the frequency and momentum dependence of the
spectral density, $A_{\bf k}(\omega)= -\frac{1}{\pi} {\rm Im} G_{\bf
  k}(\omega)$, for a typical
underdoped system.
In Fig.~\ref{fR3} we show, in the inset, the Fermi surface, defined by
those ${\bf k}$-points which fulfill
\begin{equation}
\omega=\varepsilon_{\bf k} +{\rm Re}\Sigma_{\bf k}(\omega)
\label{qp_sol}
\end{equation}
for $\omega=0$, for interacting quasiparticles whose bare spectrum is
specified by $t=-0.25 \, {\rm eV}$, $t'=-0.35t$, at a hole doping
concentration,
$n_{\rm h}=0.16$. In this and all subsequent plots we assume
$g=0.8$ eV, in agreement with transport measurements \cite{SP96}.
This corresponds to an intermediate regime for the coupling constant
since it  is similar to the total bandwidth. 
The calculation is carried out at a temperature such that $\xi=3$,
which, as we shall show, lies in the weak pseudogap regime well below
$T^{\rm cr}$.
In the main part of  Fig.~\ref{fR3}, we show our results for $A_{\bf
  k}(\omega) f(\omega)$, where $f(\omega)$ is the Fermi function
for several points on the Fermi surface.
 $A_{\bf
  k}(\omega) f(\omega)$, the quantity measured in ARPES experiments,
  is strongly anisotropic.
For a representative cold quasiparticle (a), located close to the
diagonal, with $|\varepsilon_{\bf k}-\varepsilon_{{\bf k}+{\bf Q}} | >
v/\xi$, the peak in the spectral density crosses the Fermi surface.
For these quasiparticles, the quasistatic magnetic correlations simply
produce a thermal broadening of
the spectrum, as is characteristic of a Landau Fermi liquid at small
but finite $T$.

The situation is completely different for the hot quasiparticles  at (d)
which are 
located close to
$(\pi,0)$. Here, 
$|\varepsilon_{\bf k}-\varepsilon_{{\bf k}+{\bf Q}} |< v/\xi$.
A large amount of the spectral weight is shifted to higher
energies, a shift which gives rise to 
 weak pseudogap behavior.
As will be discussed below, the position of the maximum of this broad
feature, which represents the incoherent part of the single particle
spectrum (i.e.,
does {\em not}
 correspond to a solution of
Eq.~\ref{qp_sol}), is  similar to 
   the quasiparticle bands of a mean 
field spin density wave state.
Thus, even though incoherent in nature and considerably broadened,
this high energy feature is the precursor effects of a spin density
wave state.
A second  interesting aspect of the  calculated 
 hot quasiparticle spectral density
is that although there exists 
a solution of Eq.~(\ref{qp_sol})  at $\omega=0$ those 
quasiparticles [and quite generally those near $(\pi,0)$] 
do not possess a peak. 
  This part of the FS is therefore   
{\em not} observable in an ARPES experiment. 
Experimentally, a FS crossing can only be determined if  a local 
maximum 
of the spectral density crosses the Fermi energy. The  calculated
{\em visible} part of the FS, where our calculated spectral function 
exhibits a  maximum 
at $\omega=0$, is shown in Fig.~\ref{fR4} (thick lines). 
It is in agreement with experiment.
While  this behavior appears to be similar to that expected for
a hole pocket, below we
discuss the important differences
between our results and a hole pocket scenario.

The  reason for the ``disappearance''
of pieces of the Fermi surface in the weak pseudogap regime is the
following.
The finite imaginary part of the self energy at 
$\omega=0$ invalidates, as always for $T\neq 0$, a rigorous quasiparticle
 picture and can even affect the occurrence of a maximum of the 
spectral density   in the solution of Eq.~\ref{qp_sol}.
 This is what happens for hot quasiparticles in the
weak pseudogap regime. Due to their strong magnetic
 interaction the related  large scattering rate  causes  the  hot 
 quasiparticle peak to be  invisible in the
weak pseudogap regime  and care must be taken to 
properly interpret the
calculated Fermi surface.

Consider now the evolution of the Fermi surface with temperature, or
 what is equivalent, with $\xi$. As 
 can be seen in  Fig.~\ref{fR4}, for $\xi=1$, the FS is basically unaffected
by the correlations, a situation very similar to the one obtained within
 a self consistent one loop calculation. This confirms  the result
 obtained by Monthoux,\cite{pm-vertex}
that vertex corrections, neglected in the one loop framework, are indeed
of minor importance for  small correlation lengths.
[Note that while 
our calculations are based on the fact that the dominant 
momentum transfer occurs near the antiferromagnetic wave vector, which implies
at least an intermediate correlation length $\xi$, it
is useful to consider the limiting case,  
$\xi \approx 1$, even though in this regime  different
theoretical approaches  may be turn out to be  more appropriate.]
On increasing $\xi$ to values which are realistic for underdoped cuprates 
($\xi=2 \cdots 8$), we find   slight 
changes of the FS-shape for momenta close to $(\pi,0)$ and $(0,\pi)$;
however, the general shape (large FS closed around $(\pi,\pi)$ and equivalent 
points) remains the same.
If one further increases $\xi$ to values larger then $10$ lattice 
constants, serious modifications of the FS,  
caused by a  short range order induced
flattening of the dispersion of the  quasiparticle  solution,  begin to  occur.
This  follows from the solution of  Eq.~\ref{qp_sol}   for finite
$\omega$. 
It is only for such  large correlation lengths   that
 a hole pocket starts to form  along the diagonal.
Eventually, at some large, but finite value of 
$\xi$, our solution gives a closed hole pocket.  We conclude that 
for underdoped  but still superconducting cuprates
(with $\xi \leq  8$),
 the shape
of the FS remains basically unchanged, 
while our theory can potentially describe
the transition from a large Fermi surface to a situation with a hole pocket
around $(\pi/2,\pi/2)$,  which may be 
the case  very close to the half filling.

The above results  provide a natural explanation for what is
seen at temperatures below the superconducting transition in ARPES
experiments on the underdoped cuprates: the sudden appearance of a
peak in the spectrum of quasiparticles located near
$(\pi,0)$.  According to  our results, this is to be expected, since
as $T$ falls below $T_{\rm c}$, the scattering rate
of the hot quasiparticles drops dramatically; the superconducting gap
has suppressed the strong low frequency scattering processes which
rendered invisible the peak in the normal state, and a quasiparticle
peak emerges. Since this sudden appearance of the quasiparticle peak
below $T_{\rm c}$ is inexplicable in a hole pocket scenario, the ARPES
experimental results support the large Fermi surface scenario we have
set forth above.

Another interesting aspect of the calculated results shown in
Fig.~\ref{fR3} is the sudden transition between
 hot and cold quasiparticles,  justifying  the usefulness of  
this terminology {\em a posteriori}. 
To demonstrate explicitly
the anisotropy of the spectral function for low frequencies,
we show, in Fig.~\ref{fR5},  $A_{\bf k}(\omega=0)$ along the Fermi surface as 
function of the angle $\phi_{\rm F}={\rm arctan}(k_y/k_x)$
between ${\bf k}$ and the $k_x$ axis.
Even though no gap occurs in the hot quasiparticle
spectral density in the weak pseudogap regime, the low frequency spectral
 density is considerably reduced.
It is therefore not possible to consider the behavior above the strong
pseudogap
crossover 
temperature, $T_*$, where our theory should apply, as being conventional.

We  compare,  in Fig.~\ref{fR6},  
the calculated  variation of the maximum of $A_{\bf k}(\omega)$ 
in momentum space  with  the ARPES results of
 Marshall {\em et al.}~\cite{MDL96}
for
two different doping concentrations.
For an overdoped system, we assumed a correlation length $\xi=1$ and a
charge carrier concentration $n_h=0.22$.
The resulting dispersion corresponds to that  of the original tight
binding band with slightly reduced band width. 
The plotted maxima for $\xi=1$ all correspond to broadened
coherent 
quasiparticle states. We chose $t'=-0.35t$  leading to
a Fermi surface crossing along the diagonal as well as between
$(\pi,0)$ and $(\pi,\pi)$ in agreement with experiments
The situation is different for  an underdoped system, which   we assumed
to have a charge carrier concentration $n_h=0.16 $
and a correlation length $\xi=3$, similar to other underdoped but
superconducting
cuprates. We use the same value $t'=-0.35t$ for the next nearest
neighbor hopping integral.
Along the diagonal, we still find a Fermi surface crossing and, in
agreement with experiment, no doping dependence of the Fermi
velocity of cold quasiparticles.
However, for hot quasiparticles  close to  $(\pi,0)$, only the
incoherent  high energy
feature around $200\, {\rm meV}$ is visible.
The momentum dependence of this high energy feature, even though
incoherent in its nature, is similar to 
 the  
dispersion of a mean field SDW  state:
\begin{equation}
E^{\pm}_{\bf k}=\frac{1}{2}(\varepsilon_{\bf k} + \varepsilon_{{\bf k}+{\bf Q}} )
\pm \sqrt{\left(\frac{\varepsilon_{\bf k} - 
\varepsilon_{{\bf k}+{\bf Q}} }{2}\right)^2 + \Delta_{\rm SDW}^2}\, ,
\label{sdw_disp}
\end{equation}
where  $\Delta_{\rm SDW}^2=\frac{2}{3} \Delta^2$, as can be obtained from the 
saddle point approximation of the Borel summed $\xi \rightarrow \infty$ perturbation 
series (see appendix B).
This provides an explicit demonstration 
that  the high energy feature is indeed  an
incoherent precursor of an SDW state. 
The agreement between theory and experiment regarding the
detailed momentum dependence of the high energy feature, is an
important
confirmation of the general concept of our approach.

While the overall position of the high energy feature 
($\approx 200\, {\rm meV}$ in the
present case) depends on the value of $t'$,
the general momentum dependence of these states
remains robust against any reasonable variation of $t'$ or the
coupling constant $g$. We note that 
the experiment  of  Marshall {\em et al.} ~\cite{MDL96} was performed in
the strong pseudogap state. It is however natural to expect that the
high energy
feature remains unaffected by the
opening of the low frequency leading edge gap; it    will thus  be the same in
the weak and strong pseudogap state, and will be little affected by the
superconducting transition.

In ARPES experiments at half filling, it is found  that  
 the location of momentum states      with half of the intensity of a
completely occupied state, i.e., with  $n_{\bf k}=\frac{1}{2}$,    
 is nearly    unchanged
compared to the case at large doping.
 We show
in Fig.~\ref{fR2} our calculated results for the momentum points with 
$n_{\bf k}=\frac{1}{2}$; our results are quite  similar   for  the physically
relevant values $1 \leq \xi \leq 4$ and change only slightly for large  correlation 
lengths.  Note, that in the latter case $n_{\bf k}$ varies only  
gradually. Even very  close to $(\pi,0)$   it would  be  hard  to determine
experimentally  whether
$n_{{\bf k}=(\pi,0)}$ is larger or smaller than one half.
 Our  results are therefore in   agreement with the experimental
situation; they
demonstrate that there is a "memory" in the correlated
system which, as far as the total charge of a given ${\bf k}$-state
is concerned,  behaves quite  similarly  to the 
case without strong antiferromagnetic
correlations.

Finally, we   address the question of why, for   moderate values
of the
correlation length, we   obtain such pronounced
anomalies.
In addition to $\xi$, 
the only length scale  in the problem is the electronic length
$\xi_0\approx v/\Delta \approx 2  v/g$. It is natural to 
argue  that once
$\xi > \xi_0$ some  new behavior of the quasiparticles
 due to short range order might appear.
Within standard weak coupling theories, $\xi_0$ is by construction a large quantity, and the theory
is trustworthy only for very large $\xi$. The summation of the entire perturbation series
in our calculation however enables us to take  account for the situation 
where $\xi_0$ can be of the order of a few lattice constants, i.e. for the
intermediate coupling constant regime.
That this qualitative argument is also quantitatively correct, can be seen in
 Fig.~\ref{fR7}, where we show the $\xi$ dependence of the spectral
density at a
hot spot for which $\varepsilon_{\bf k}=\varepsilon_{{\bf k}+{\bf Q}}$.
For the above given set of parameters, $\xi_0 \approx 2$ and 
SDW precursors occur as soon as
$\xi > \xi_0$. This is in striking agreement with, and provides a
microscopic explanation for,  the  prediction
 by Barzykin and Pines that one finds $\xi(T=T^{\rm cr}) \approx 2$
at the crossover  temperature $T^{\rm cr}$, where the
 magnetic response changes character.

We conclude 
that the quasiparticle excitations 
in the weak pseudogap regime 
 are intermediate between
a conventional system with a large Fermi surface and a spin density wave
system with a
 small Fermi surface.
The fact that both aspects are relevant  explains  the failure of any approach which
concentrates on only one of these.

\section{spin and charge vertex functions}
  We turn now to  the coupling 
of quasiparticles of the weak pseudogap state 
with the collective spin and 
charge degrees of freedom. This is of interest in  its own right, and
is of importance
for an understanding of the charge and spin response functions
discussed in II.
The quantities which characterize the interaction of quasiparticles with the   spin and 
charge degrees of freedom are the vertex functions 
$\Gamma^{ {\rm s}}_{{\bf k},{\bf k}+{\bf q}}
(\omega+i0^+,\omega+\nu+i0^+) $ and  $\Gamma^{ {\rm c}}_{{\bf k},{\bf k}+{\bf q}}
(\omega+i0^+,\omega+\nu+i0^+) $.
In order to have an idea of the   behavior of these vertex functions we 
first consider their    behavior analytically  in  the lowest nontrivial
 order of the perturbation series. Our subsequent numerical
 results are 
obtained from the full solution of the problem.
For the spin vertex, we find on using Eq.~\ref{vertex} and  
 Eq.~\ref{momapp} for $S(q)$, that up to second order in $g \propto \Delta$:
\begin{eqnarray}
\Gamma^{ {\rm s}}_{{\bf k},{\bf k}+{\bf q}}
(\omega+i0^+, \omega+\nu+i0^+)&=& 1-\frac{\Delta^2}{3} \frac{1}
{\omega-\varepsilon_{\bf k}
+i v/\xi} \nonumber \\
& &\frac{1}{\omega+\nu -\varepsilon_{{\bf k}+{\bf Q}}
+i v/\xi } 
\label{ver_seco}
\end{eqnarray}
We are mostly interested in the vertex function for frequencies $\omega$ which
correspond to the quasiparticle energies at the Fermi surface.
 For the case of an unchanged Fermi surface,  the bare dispersion
$\varepsilon_{\bf k}$   determines the quasiparticle energies 
at this Fermi surface.
Once hole pockets are formed, these are given by the SDW energies $E^{\pm}_{\bf k}$ of
Eq.~\ref{sdw_disp}.
On evaluating  Eq.~\ref{ver_seco}   at the SDW energies $\omega= E^{\pm}_{\bf k}$ of
Eq.~\ref{sdw_disp} and for $\nu=0$ in the limit 
$\xi \rightarrow \infty$, we find, 
  $\Gamma^{ {\rm s}}=\frac{2}{3}$;   the spin vertex is reduced.
For the case of long range antiferromagnetic order, with only two spin degrees of freedom left,
$\Gamma^{ {\rm s}}= 0$ vanishes, as was shown by Schrieffer~\cite{Bob}.
On the other hand, if one  takes into account that in the weak
pseudogap regime the Fermi surface 
is basically unchanged  and evaluates Eq.~\ref{ver_seco}
at small frequencies $\omega=\nu=0$   for a hot spot with 
$\varepsilon_{\bf k}=\varepsilon_{{\bf k}+{\bf Q}}=0$, it follows  that
\begin{equation}
\Gamma^{ {\rm s}}=1+\frac{\Delta^2}{3 v^2} \xi^2\, ,
\end{equation}
  i.e. the vertex is considerably {\em enhanced}.
In the case of the charge vertex the prefactor $\frac{1}{3}$ in Eq.~\ref{ver_seco}
has to be replaced by $-1$ and one finds $\Gamma^{ {\rm c}}=4$ if one considers
$\xi \rightarrow \infty$ and $\Gamma^{ {\rm c}}=1-\frac{\Delta^2}{ v^2} \xi^2$
in the case of an unchanged Fermi surface.
These considerations demonstrate that  which energies one considers
and how the Fermi surface evolves is crucial for an understanding of the role of vertex
corrections, i.e., enhancement vs. suppression. It also shows that only
a careful and 
self consistent analysis can reveal
in which  way the renormalized charge and spin interactions vary. This
we now do.

In Fig.~\ref{fR8} we show the spin vertex $\Gamma^{ {\rm s}}_{{\bf k},{\bf k}+{\bf Q}}
(\omega+i0^+,\omega+i0^+) $ for hot and cold quasiparticles with momentum transfer 
${\bf Q}$ and zero frequency transfer as function of energy and, for
comparison, 
  the corresponding
spectral function,    $A_{\bf k}(\omega)$. 
As in the case for the spectral function, the vertex function is
strongly 
anisotropic;  for cold quasiparticles  vertex corrections are negligible, whereas
the strong low frequency enhancement of $\Gamma^{ {\rm s}}_{{\bf k},{\bf k}+{\bf Q}}
(\omega+i0^+,\omega+i0^+) $ for hot quasiparticles demonstrates that  despite their
 reduced 
low frequency spectral weight, hot quasiparticles interact
 strongly with the spin fluctuations.
Thus, for physically reasonable $\xi$-values, 
the low frequency vertex is not reduced but enhanced.
This is  important  and must be taken  into account 
in constructing an effective theory for
the low energy
 degrees of freedom of the strong pseudogap
state. Even without detailed calculations, it is evident that 
in the spin fluctuation model 
the strong coupling 
nature of the low frequency degrees of freedom is 
 crucial since it  demonstrates that quasiparticles and spin 
fluctuations do not decouple.
This is essential to obtain a spin fluctuation induced superconducting state and
suggests that  new  strong coupling phenomena  are likely to occur once
 the temperature decreases and the system changes character due to the suppression
of the quasi particle scattering rate.

We further note that these results are frequency dependent. Thus when we
 consider the vertex at frequencies  $\omega$ close to 
the high energy features,
we find a moderate reduction, which has the same  origin   as   the
vanishing 
vertex of the long range ordered state~\cite{Bob}.

The anisotropy of the spin vertex function can   be seen in Fig.~\ref{fR9}, 
where we plot the spin vertex $ \Gamma^{ {\rm s}}_{{\bf k},{\bf k}+{\bf Q}}
(\omega+i0^+,\omega+i0^+) $ for $\omega=0$ along the Fermi surface 
as a  function of 
$\phi_{\rm F}= {\rm arctan}(k_y / k_x)$ (solid line).
The  rather sharp  transition between hot and cold quasiparticle
behavior is similar to that seen for  the  spectral density,  in Fig.~\ref{fR5}.

The corresponding behavior for the charge vertex function is shown in
Fig.~\ref{fR10}, which 
    shows that 
$ \Gamma^{ {\rm c}}_{{\bf k},{\bf k}+{\bf q}}
(\omega+i0^+,\omega+i0^+) $ behaves in  opposite fashion
to the spin vertex.
{\em At low frequencies, the quasiparticles are almost 
completely decoupled from
potential collective  charge degrees of freedom}. This effect is strongest for 
hot quasiparticles,  and occurs for  
momenta transfers ${\bf q}$,  around $(0,0)$ as well as those close to ${\bf Q}$. 
The formation of incoherent spin density wave precursors obviously leads to a decoupling of
the low energy quasiparticles from the  charge degrees of freedom.
 In our theory, its origin is the dominant interaction of the
 quasiparticles  with collective
spin degrees of freedom.
It  is thus   impossible that any kind of static or low frequency dynamical
charge excitation can substantially  affect the charge carrier
dynamics of  hot quasiparticle states in cuprates.
Furthermore, this result explains another important puzzle of the
cuprates: 
the irrelevance to transport phenomena in the normal state
of the total electron phonon coupling,
\begin{equation}
g^{\rm el.-ph.}_{{\bf k}, {\bf q}} \approx g^{{\rm el.-ph.}(0)}_{{\bf k}, {\bf q}}
\Gamma^{ {\rm c}}_{{\bf k},{\bf k}+{\bf q}}
( 0+i0^+,0+i0^+),
\end{equation}
despite their pronounced ionic structure, which
in fact suggests a strong bare  interaction $g^{{\rm el.-ph.}(0)}_{{\bf k}, {\bf q}} $ of
charge carriers with the poorly screened lattice vibrations.

These considerations show that an effective low energy theory of
the weak pseudogap state must take the  strong and anisotropic spin fermion interaction 
into account; it    can safely neglect the coupling to phonons of hot
quasiparticles  as well as any other 
interaction they might have 
with charge excitations. This is an important new theoretical constraint 
for the strong pseudogap state.

\section{Conclusions}
  We have used our solution of the quasistatic
spin fermion model of  antiferromagnetically correlated spin
fluctuations to  develop a new  description of    the intermediate  weak
pseudogap state of underdoped cuprate superconductors.
 Based on  the experimental observation that the characteristic energy
scale of over-damped spin excitations is small compared to the
temperature, once $T$ lies between the two crossover scales $T_*$ and
$T^{\rm cr}$, we conclude that the spin degrees of freedom behave
quasi-statically
i.e. the spin system is thermally excited and the scattering of
quasiparticles
with their own collective spin modes can be regarded as resulting from
a static spin
deformation potential,   characterized only by the strength and spatial
extent of these spin fluctuations. 
On neglecting the quantum dynamics of the spin modes, we were able to 
solve this spin fermion
model by summing all diagrams of the perturbation series for the
single
particle Green's function and the spin and charge vertex functions.
This enabled us to directly investigate the spectral function,  
measured in angular resolved photo-emission experiments,    the effective
interactions of quasiparticles with spin as well as charge collective modes
and the spin and charge response behavior.

Our results demonstrate  that for intermediate values of the
antiferromagnetic correlation length and intermediate coupling
constant $g$, one is dealing with the 
rich physics  of a crossover between   non-magnetic
and   spin density wave like behavior. While the Fermi surface of
the quasiparticles remains unchanged their highly anisotropic 
effective interaction
leads to  two different classes of quasiparticles, {\em hot} and {\em
  cold},  with  only the   hot quasiparticles feeling  the full
strength of
the antiferromagnetic interaction.  For the latter,  a transfer of
spectral weight to  high energy features occurs. These high
energy features are the  incoherent precursors of a spin density wave
state;  their momentum dependence is in excellent agreement with
corresponding high energy structures at around $200\, {\rm meV}$ seen
in recent ARPES experiments. 
For low energies, hot quasiparticles have a reduced spectral weight,
a {weak pseudogap characteristic}, and the coherent quasiparticle
poles are completely over-damped due to the strong scattering rate.
The drop in  scattering rate found in the   strong pseudogap state
is not sufficient to make this  
 quasiparticle pole    visible. It is only below $T_c$ that the rate drops
 sufficiently that the pole becomes visible.
This scenario explains
the appearance of a sharp peak, invisible
above $T_c$,  at low frequencies in the
superconducting state of underdoped cuprates.
The high energy features on the other hand are expected
to be unchanged as temperature is lowered.

Finally, we used our calculation of the irreducible
vertex functions to
 investigate the effective interaction of the
quasiparticles with spin and charge modes.
We find that this effective interaction is likewise highly
anisotropic;
the low energy electron-spin fluctuation interaction is strongly
enhanced whereas the coupling to charge degrees of freedom is reduced.
The  enhancement of the spin vertex is essential  for  
the development of a  spin
fluctuation induced  superconducting state and is an indicator as well of 
anomalous behavior  at lower temperatures. The reduction of the
charge
vertex causes a reduced electron-phonon coupling constant for  hot
quasiparticles as well as a decoupling of hot quasiparticles from
potential  charge collective modes.

This scenario applies, as discussed, for temperatures between $T_*$
and $T^{\rm cr}$. Below $T_*$, the characteristic frequency of the
spin system increases for decreasing temperature,  making  it
impossible to consider the system as exhibiting quasistatic behavior.
Thus,
 here we expect that the quantum nature of the spin degrees of
freedom becomes essential; it brings about a strongly reduced phase space for
the inelastic scattering of quasiparticles and spin
fluctuations, and
causes a sudden drop in the corresponding scattering rate of
the system. Nevertheless, since we expect the high energy features to
remain unchanged, the spectral weight of hot  quasiparticles is
strongly
reduced while their interaction is enhanced. This strong coupling
behavior can   bring about precursor to superconducting state behavior, as
recently discussed by Chubukov~\cite{Cub97}, who  
found that at  zero temperature, the quantum behavior of the
spin modes   causes  a leading edge gap of the spectral function
within
a magnetic scenario for cuprates. It has recently been shown that this
leading edge gap will be filled with states once the temperature is
raised, 
and that it  basically disappears for $T \approx 0.6 \omega_{\rm sf} \xi^{1/
  \nu}$, with $\nu \approx 2$~\cite{CS98}.   Above this temperature,
 the system starts to
behave quasistatically and the  superconducting precursors  are   suppressed.
 Within the same scenario, one thus finds that  the 
classical behavior of spin fluctuations favors magnetic precursors
and
causes an increase of the low energy interaction,  which leads 
for lower temperatures 
 to  superconducting precursors due to the quantum behavior of the
 spin fluctuations. This provides perhaps the   first consistent picture of the
 microscopic origin of the two pseudogap regimes and their crossover
 temperatures:
The strong pseudogap state and the leading 
 edge gap are the quantum manifestation of strong
antiferromagnetic correlations whereas the spin density wave
precursors
are its  classical 
manifestation.

What are the consequences of this 
new microscopic scenario for  the  crossover behavior 
  of underdoped cuprates?
Our theory predicts that  upon measuring the uniform
susceptibility $\chi_o(T)$ and the spectral function for an slightly
underdoped or optimally doped system, (for which  $T^{\rm cr}$ is not too
high),  it will turn out that above 
 $T^{\rm cr}$, where $\chi_o(T)$ is maximal, the high energy features
 will disappear.
We also expect important insights  into the role of impurities and high
magnetic fields.
Non-magnetic impurities  should  affect the weak pseudogap state only
slightly. Their strongest effect will occur slightly below $T^{\rm cr}$
where    impurities cause $\xi$ to   decrease to $\xi_{\rm imp}$.
Provided $\xi_{\rm imp} < \xi_o\approx 2 < \xi $, an impurity
driven transition out of the weak pseudogap state occurs.
$T_*$ will be considerably more sensitive to impurities because
for particle-particle excitations with a  tendency to d-wave pairing,
non-magnetic impurities act destructively  and the strong pseudogap 
state may even completely disappear.

High magnetic fields provide another  indicator of  the
differences
between strong and weak pseudogap behavior.
In the weak pseudogap regime no sensible effects for all achievable
field strengths will occur, because the relevant coupling of the
magnetic field is the Zeeman interaction with the spins, $\propto {\bf
  H}\cdot {\bf s}$,  which has to compete with the much stronger short
range correlations of these spins.
In the strong pseudogap state,   we expect  that the dephasing due to
the 
minimal coupling ${\bf p} \rightarrow {\bf p}-\frac{e}{c} {\bf A}$
will strongly affect the behavior in the pairing channel, causing a
suppression of   strong pseudogap  behavior.
From this perspective, it immediately follows  that the transport
experiments by Ando {\em et al.}~\cite{ABP95} in a pulsed high magnetic field 
 demonstrate that weak pseudogap behavior is prolonged to lower
 temperatures.
It is tempting to speculate that for very  strong  magnetic fields the
weak pseudogap regime crosses directly over into an insulating one.

In general , for decreasing temperature the weak pseudogap regime 
can therefore  cross over to a strong pseudogap state or to an insulator
and,  under certain circumstances,  directly into the superconducting
state.  The latter may occur for systems with very large
incommensuration, which tend to reduce the scattering rate of
a quasistatic spin system.~\cite{note_incom};  it   is likely  of
relevance  for  La$_{2-x}$Sr$_x$CuO$_4$.

 Our results for the low frequency spin dynamics 
and optical response as well as the Raman intensity in the B$_{1g}$
channel will be compared    with experiment
 in a subsequent paper.
  There we  also  demonstrate   that we can explain the 
generic changes of the low frequency
magnetic response at the upper
 crossover temperature $T^{\rm cr}$~\cite{SPS97}.
 
\section{Acknowledgments}
This work has been supported in part by the Science and Technology
Center for Superconductivity through NSF-grant DMR91-20000, 
by the Center for Nonlinear Studies and Center for Materials Science
at Los Alamos National Laboratory,
and by the Deutsche Forschungsgemeinschaft (J.S.).
 It is our pleasure to thank M. Axenovich, G. Blumberg,
 J. C. Campuzano,  A. Chubukov,  D. Dessau,  O. Fischer,
A. Millis,  D. Morr,  Ch. Renner, Z.-X. Shen, C. P. Slichter,  
R. Stern and O. Tchernyshyov for  helpful discussions.

\appendix
\section{Single Particle Green's Function and Diagrammatic Rules}
\label{App_GF}
In this appendix we derive the diagrammatic rules of the spin fermion model
 under circumstances that  the interaction between collective spin modes is neglected.
Note, these diagrammatic rules do not rely on the quasi-static
approximation, but are completely general for the spin
fermion model.

Using the generating functional
\begin{equation}
W[\eta,\eta^\dagger]= \frac{1}{Z} \int {\cal D}c^\dagger {\cal D}c \,
\exp\left\{-(S+ c^\dagger \eta +\eta^\dagger c)\right\}
\label{genfun}
\end{equation}
 with partition function $Z=\int {\cal D}c^\dagger {\cal D}c \,
 e^{-S}$,
effective action of Eq.~\ref{act_spinf}
and short hand notation
  $c^\dagger \eta=\int_0^\beta d\tau \sum_{{\bf k} \sigma } 
c^\dagger_{{\bf k}\sigma}
\eta_{{\bf k},\sigma}$,
the single particle Green's function can be obtained via  a functional
derivative:
\begin{equation}
G_{{\bf k},\sigma} (\tau - \tau') 
 =\left. \frac{\delta^2 W[\eta, \eta^\dagger]}{\delta \eta_{{\bf k} \sigma}^\dagger(\tau)
\delta \eta_{{\bf k} \sigma}(\tau')} \right|_{\eta=\eta^\dagger=0} \, .
\end{equation}
As usual $c$, $\eta^\dagger$ etc. are Grassman variables.
In order to perform this functional derivative  it is convenient  to  introduce 
 a collective bosonic spin 1 field ${\bf S}_{\bf q}$
by adding an irrelevant Gaussian term to the action $S \rightarrow S+S_o$, where
\begin{equation}
S_o({\bf S})=\frac{1}{2} \int_0^\beta d\tau \int_0^\beta d\tau'  
\sum_{{\bf q}} \chi_{\bf q}^{-1}(\tau -\tau') \, {\bf S}_{\bf q}(\tau) \cdot {\bf S}_{-{\bf q}} (\tau')\, ,
\label{coll0}
\end{equation}
to integrate with respect to this spin field and to divide by the corresponding partition function
$Z_{\rm B}$ of this ideal Bose gas. Finally, we shift (after Fourier transformation   from time
to frequency) the variable of integration as:
\begin{equation}
{\bf S}_{\bf q}(i\omega_n) \rightarrow {\bf S}_{\bf q} (i\omega_n) -\frac{2 g}{\sqrt{3}}  
\chi_{\bf q}(i\omega_n)  {\bf s}_{\bf q} (i\omega_n)  \, , 
\end{equation}
leading to the effective action of the {\em spin fermion model}:
\begin{eqnarray}
S &=&-\int_0^\beta d\tau \sum_{{\bf k},\sigma} c^\dagger_{{\bf k}\sigma} 
G^{-1}_{o {\bf k}} c_{{\bf k}\sigma} 
+ S_o
\nonumber \\
& &+\frac{2g}{\sqrt{3} }  \int_0^\beta d\tau \sum_{{\bf q}} 
{\bf s}_{\bf q}(\tau) \cdot {\bf S}_{-{\bf q}} (\tau)\, .
\end{eqnarray}
After this Hubbard Stratonovich transformation,   
 we can integrate out the fermions, yielding
\begin{equation}
W[\eta,\eta^\dagger]= \frac{1}{Z Z_{\rm B}} \int {\cal D}{\bf S}
\exp\left\{-(S_c({\bf S}) -\eta^\dagger \hat{G}({\bf S}) \eta)\right\} \, ,
\end{equation}
with  the   action of the collective spin degrees of freedom
\begin{equation}
S_c({\bf S})=-{\rm tr} \, {\rm ln}\left( -\hat{G}({\bf S} )^{-1} \right) + S_o({\bf S})\, .
 \end{equation}
Here,  we have introduced the matrix Green's function
\begin{eqnarray}
\hat{G}^{-1}_{{\bf k},{\bf k'}}(\tau,\tau' |{\bf S} )&=&
G^{-1}_{o{\bf k} }(\tau-\tau' ) \delta_{{\bf k},{\bf k'}} \sigma_o \nonumber \\
& &-\frac{g}{\sqrt{3}} {\bf S}_{{\bf k}-{\bf k'}}(\tau)\delta(\tau-\tau')
\cdot {\bf \sigma}\, ,
\label{GFmatrix}
\end{eqnarray}
which describes the propagation of an electron for a given configuration ${\bf S} $ of the spin field.
Performing the above functional derivative with respect to 
$\eta$ and $\eta^\dagger$ gives
finally
\begin{eqnarray}
G_{{\bf k} \sigma}(\tau-\tau')&= &\frac{1}{Z} \left\langle
 \hat{G}_{{\bf k},{\bf k} \sigma \sigma }(\tau,\tau' |{\bf S} ) \right. \nonumber \\
& &\left. \exp\left\{ {\rm tr} \, {\rm ln}\left( -\hat{G}({\bf S} )^{-1} \right) \right\} \right\rangle_o ,
\label{GF}
\end{eqnarray}
where the average 
\begin{equation}
\langle \cdots \rangle_o= \frac{1}{Z_{\rm B}} \int {\cal D}{\bf S} \cdots
\exp\left\{-S_o\right\}
\label{aver_o}
\end{equation}
is performed  with respect to the free collective action of Eq.~\ref{coll0}.
This is   a standard  exact reformulation  of  Eq.~\ref{act_spinf}
of collective spin fields which has the appeal that   explicit fermion degrees
no longer occur and that Wick's theorem  of the Bose field ${\bf S}$ 
can be used to  evaluate the single particle Green's function.

Since we do not expect the interaction of the spin modes to be relevant, we
neglect  nonlinear (higher order in ${\bf S}$ than quadratic)  terms
of the spin field,  assuming that no modifications due to
 spin fluctuation-spin fluctuation interactions occur  beyond  those 
already included in $\chi_{\bf q}(\omega)$.
Using renormalization group arguments, it has been shown by Millis~\cite{andy}
that indeed the system is characterized by a Gaussian fixed point as far as the
low frequency spin dynamics is concerned.
The  mathematical  consequence of these two assumptions
 is that we can use the approximation
\begin{equation}
\exp\left\{ {\rm tr} \, {\rm ln}\left( -\hat{G}({\bf S})^{-1} \right)
\right\}  \approx  Z\, .
\label{cons_simpl}
\end{equation}
Contributions of second order in ${\bf S}$  can be ignored,
since they would renormalize $\chi_{\bf q}(\omega)$, which
is   assumed to be the experimentally determined,
i.e. fully renormalized susceptibility.
Eq. ~\ref{cons_simpl}
leads to  a considerably   simplified expression
 for the Greens function:
\begin{equation}
G_{{\bf k} \sigma}(\tau-\tau')=   \left\langle
 \hat{G}_{{\bf k},{\bf k} \sigma \sigma }(\tau,\tau' |{\bf S} ) \right\rangle_o \, .
\label{GFapprox}
\end{equation}

Consequently,  the diagrammatic series for the determination of the single particle Green's
function reduces to that  of a single particle problem with time dependent spin "impurities".
Inversion of the Greens function matrix of  Eq.~\ref{GFmatrix} in spin space yields for  
 the  $\sigma$-spin matrix-element:
\begin{eqnarray}
G_\sigma ({\bf S})&=&\left( G_o^{-1}- \frac{\sigma g}{\sqrt{3}}S^z- \right.
\nonumber \\
& & \left.\frac{g^2}{3} S^{-\sigma} \left(  G_o^{-1}+ 
\frac{\sigma g}{\sqrt{3}}S^z \right)^{-1} S^{\sigma} \right)^{-1} \, ,
\label{invers_spin}
\end{eqnarray}
with $S^{\sigma}=S^{\pm}$ if $\sigma=\pm 1$.
Here, we still have to take the matrix nature of $G_o$ and ${\bf S}$ (in momentum
and frequency space) into account since they
are diagonal in different representations ($G_o$ in momentum and frequency space, 
${\bf S}$ in coordinate  and time space).
Considering the limiting case of only one longitudinal spin mode generated by $S^z$, 
gives after averaging:
\begin{equation}
\left\langle G_l({\bf S}) \right\rangle_o =G_o \sum_{N=0}^\infty \left( \frac{g^2}{3} \right)^N\,
\left\langle \left( S^z G_o S^z G_o \right)^N\right\rangle_o
\end{equation}
where we used the fact that odd  orders in $S^z$  vanish  without global symmetry breaking.
Using this representation of the Greens function,
the diagrammatic rules which correspond to the averaging with respect to $ S^z$
follow straightforwardly.
The averages can be evaluated  via contractions based on Wick's theorem using 
$\langle S^z_{\bf q} (\tau)  S^z_{-{\bf q}} (\tau')  \rangle_o= \chi_{\bf q}(\tau-\tau')$.
This leads to the
  diagrams of a theory of fermions interacting with a scalar  time
dependent field. 
Here, the topology of each diagram is identical
to the topology of the contraction symbols which occur by applying  Wick's 
theorem. 
Alternatively, one can also consider the case of   two  transverse modes leading to
\begin{equation}
\left\langle G_t({\bf S}) \right\rangle_o =G_o \sum_{N=0}^\infty \left( \frac{g^2}{3} \right)^N\,
\left\langle \left( S^- G_o S^+ G_o \right)^N\right\rangle_o
\end{equation}
which is identical to a theory of fermions interacting with a
"charged" time dependent  field. 
Again Wick's decompositions using $\langle S^+_{\bf q} (\tau)  S^-_{-{\bf q}} (\tau')  \rangle_o= 
2 \chi_{\bf q}(\tau-\tau')$ can be  performed.  

The situation becomes more complicated  if one considers   simultaneously  
longitudinal and transverse modes.
Here, it follows from Eq.~\ref{invers_spin}:
\begin{eqnarray}
G_\sigma({\bf S})&=&G_{o,\sigma}(  S^z)\,  \sum_{N=0}^\infty \left( \frac{g^2}{3} \right)^N\,
\nonumber \\
& & \times   \left( S^- \,   G_{o, -\sigma}(  S^z) \, S^+\,  G_{o,\sigma}(  S^z)  \right)^N  \, ,
\label{gfull}
 \end{eqnarray}  
where 
\begin{eqnarray}
G_{o,\sigma}(  S^z) 
 =G_o  \sum_{N=0}^\infty  (\sigma)^N \left(\frac{g}{\sqrt{3}} S^z G_o \right)^N \, .
\label{gfhlp}
\end{eqnarray} 
Eq.~\ref{gfull} and \ref{gfhlp} mean that for   any
  transverse (spin flip) scattering event  all possible  longitudinal
  (spin conserving)
 processes   occur.
The complication of this inserted partial summation is the occurrence of the sign factor
$(\sigma)^N$ of down spins.
The $( -1)^N$ factor occurs, if there are contractions out of longitudinal  processes for
an odd number of $S^z$ fields of a $\sigma=\downarrow=-1$ Green's function $G_{o,\downarrow }(  S^z)$.
If  a  $S^z$ is  paired with  another $S^z$ of the same 
 $G_{o,\downarrow }(  S^z)$, they occur in an even number without modifying the sign.
The $(-1)^N$  enters   only   if a longitudinal field $S^z$ is contracted with 
  one which refers
to a   Greens function $G_{o,\uparrow }(  S^z)$. 
 In order to reach   $G_{o,\uparrow }(  S^z)$,  the corresponding line of the $S^z$
contraction has to cross an odd number of transverse contractions.
 From these considerations, the following diagrammatic rules for an arbitrary diagram
of order $2N$  result:
\begin{itemize}
\item  Draw  $2N+1$ solid lines with $2N$ vertices which can be $2L$
   spin conserving, 
$N-L$ spin lowering 
and $N-L$ spin raising vertices , referring to vertices of 
longitudinal processes ($S^z$), leaving transverse
processes ($S^-$) and entering transverse processes ($S^+$), respectively.
For the transverse vertices one has to
ensure that two subsequent spin raising (lowering) vertices  are separated by
one spin lowering (raising) and  an arbitrary  number of spin conserving vertices.
\item Connect the spin lowering vertices pairwise with spin
   raising ones by a wiggly line
\item Connect the spin conserving vertices pairwise with   dashed lines
\item Insert for any solid line a Green's function $G_{o{\bf
      k}}(i\omega_n)$, 
for any vertex, $g/\sqrt{3}$,
for any wiggly or dashed line $ \chi_{\bf q}(i\omega_n)$ and take 
  momentum and energy conservation at the vertices into
account.
\item Multiply with  a prefactor $2^{N-L}$  which accounts for  
the two transverse spin
  degrees of freedom ($N-L$ is the number
  of wiggly lines).
\item  Multiply with  a prefactor $(-1)^C$ in front of the diagram, 
where $C$ is the number of crossings of
dashed and wiggly lines.
\end{itemize}
Finally one has to sum over all  possible diagrams generated by this
procedure.
In Fig.~\ref{fAp1} we plot all diagrams for the self energy up to order
$g^4$, including signs and multiplicities.
The occurrence of the additional crossing sign which results from the interference
of longitudinal and transverse modes certainly complicates the situation.  
If there were only one or two components of the SU(2) spin vector, the diagrammatic rules reduce to the special
case of only  longitudinal or  transverse  interaction lines. Here, the problem    is identical to that of 
uncharged or charged bosons, respectively.  Only the simultaneous  consideration of both phenomena, which is
necessary to preserve spin rotation invariance,  leads to the prefactor $(-1)^C$ and  
reflects the fact 
 that the collective mode  of the system is a {\em spin } fluctuation.

Identical diagrammatic rules can be derived for the spin fluctuation
vertex function. In Fig.~\ref{fAp2} we show, as an example, the spin flip
vertex $\Gamma_{\rm t}$ as well as the spin conserving vertex
$\Gamma_{\rm l}$. Even though our approach is not constructed to be
manifestly 
spin rotation invariant, this symmetry must of course be fulfilled
once the transverse and longitudinal spin susceptibilities, represented
by the wiggly and dashed lines of the above diagrammatic  rules,
 are the same.
As can be seen from the lowest order vertex corrections, the above
derived diagrammatic rules guarantee indeed that $\Gamma_{\rm
  l}=\Gamma_{\rm t}\equiv \Gamma^s$, as expected. It can be shown
order by order in the coupling constant $g$ that our procedure
guarantees spin rotation invariance, as is essential for a
system with isotropic spin fluctuations.

\section{Determination of the Multiplicity of Diagrams}
\label{App_mult} 
 
In this appendix we calculate the   multiplicity of identical diagrams
for a given order of the perturbation series.
This   will be done in two steps.
First, we  solve the problem in the limit $\xi \rightarrow \infty$;  second, 
we use the general expression of Eq.~\ref{sigarb}, valid for an arbitrary diagram and finite $\xi $
and determine the missing $\xi $-independent multiplicity factors from the $\xi \rightarrow \infty$
solution. For the special cases of only longitudinal or transverse modes, our solution is the same
as Sadovskii's~\cite{Sad79}.
 It is important to notice, that 
without the result of Eq.~\ref{sigarb} it would not be possible
to determine uniquely the diagram multiplicity from the infinite $\xi$
limit. 

\subsubsection{Solution for $\xi \rightarrow \infty$}
The limit 
 $\xi \rightarrow \infty$ is not free of complications:
First, we expect that in this limit the longitudinal and transverse
spin degrees behave differently. We can ignore this problem here
  because we are only interested in  the spin rotation invariant
situation for finite $\xi$  and use the limit only for the
 mathematical  purpose of 
determining diagram multiplicities. Thus we assume spin rotation
invariance also for $\xi \rightarrow \infty$.
Second, the local moment of the susceptibility in Eq.~\ref{MMP}
diverges in the static limit logarithmically for  $\xi \rightarrow
 \infty$. 
This problem can also be avoided, because the use of Eq.~\ref{momapp}
avoids this divergence, but does not change the multiplicity of the
diagrams.
Third, the rather straightforward result that for a given order in $g$
each diagram    for $\xi \rightarrow \infty$ is  besides sign and
multiplicity identical,  leads to the following perturbation expansion
of the Green's function:
\begin{equation}
G_{\bf k}(\omega)=G_{o {\bf k}}\sum_{n=0}^\infty \, (2n+1)!!
\left(\frac{\Delta^2}{3}\right)^n\,
H_{\bf k}^n\, ,
\end{equation}
where    $H_{\bf k}=G_{o {\bf k}}G_{o {\bf k}+{\bf Q}}$,
which is in fact a divergent series. One can  then obtain
a convergent result using Borel summation of this series.
This is however not the most transparent way to solve this problem; we
choose  an alternative approach,  using the path integral representation
of the Green's function derived in appendix A, which of course gives
the same result.

In the limit of infinite antiferromagnetic correlation length,  the
 only relevant spin configuration is:
\begin{equation}
{\bf S}_{{\bf k}-{\bf k}'}= {\bf S} \, \delta_{{\bf k},{\bf k}'+{\bf Q}}
\label{Sinf}
\end{equation}
and the path integral of Eq.~\ref{aver_o} simplifies considerably to
\begin{equation}
\langle \cdots \rangle_o = \frac{1}{Z_{\rm B}} \int d^{\cal N} {\bf S} \cdots e^{-\frac{g^2 {\bf S}^2}{2 \Delta^2}} \, .
\label{average}
\end{equation}
Here, the spin rotation invariant  case  of present physical 
interest is ${\cal N}=3$. If ${\cal N}=1$ or $2$
the system consists only of longitudinal or transverse modes, respectively.
In the following we solve the problem for all three situations.
In doing so, we have to replace in all expressions $g^2/3$ by $g^2/{\cal N}$.
For example  one has to generalize the expression 
\begin{equation}
\Delta^2=\frac{1}{{\cal N} }g^2 \langle {\bf S}^2\rangle
\end{equation}
for the gap energy etc.  
Here, the partition sum of the bosons is given by
 $Z_{\rm B}=\frac{1}{2} r_{\cal N} \Gamma({\cal N}/2)
(\frac{2 \Delta^2}{g^2})^{{\cal N}/2}$ with $r_1=2$, $r_2=2\pi$ and $r_3=4 \pi$, respectively.
For ${\cal N}=1$ and $2$, the analytical inversion of the Dyson equation is evident, 
 for ${\cal N}=3$ it follows:
\begin{eqnarray}
G_{ \sigma}({\bf S})_{{\bf k},{\bf k}'}= \frac{G_{o {\bf k}}} {1-\frac{(g{\bf S})^2}{3}  H_{\bf k}} \delta_{{\bf k},{\bf k}'}
+\sigma \frac{ \frac{gS^z}{\sqrt{3}}  H_{\bf k} }{1-\frac{(g{\bf S})^2}{3}  H_{\bf k}} \delta_{{\bf k},{\bf k}'+Q} \, .
\nonumber 
\end{eqnarray}
The second, non-spin rotation invariant,  term   vanishes after
averaging,  Eq.\ (\ref{average}),
and we can finally write:
\begin{equation}
G_{\bf k}(\omega)= \left\langle 
\frac{1}{\omega -\varepsilon_{{\bf k}}-\frac{(gS)^2/3 }{\omega -\varepsilon_{{\bf k}+{\bf Q}} }} \right\rangle_o\, .
\end{equation}
 It follows that the full Green's function,  obtained in the limit
$\xi \rightarrow \infty$, is an averaged second order Green's function with fluctuating 
SDW gap $\tilde{\Delta}=\sqrt{(gS)^2/3}$. For arbitrary  ${\cal N}$, the same 
 result occurs  if one replaces the gap by  $\tilde{\Delta}=\sqrt{(gS)^2/{\cal N}}$. 
Performing the angular integration of the vector ${\bf S}$, the integral of Eq.~\ref{average}
  can   be written as:
\begin{eqnarray}
G_{\bf k}(\omega)&=& \frac{ r_{\cal N} }{ 2  Z_{\rm B} } \left( \frac{\sqrt{{\cal N}} }{  g} \right)^{\cal N}
\int_{-\infty}^{\infty} d \tilde{\Delta}  \, |\tilde{\Delta}|^{{\cal N}-1}
\nonumber \\ & &
e^{-  \frac{{\cal N}}{2} \left( \frac{ \tilde{\Delta}}{ \Delta } \right)^2} \, \frac{1}
{\omega -\varepsilon_{{\bf k}}-\frac{\tilde{\Delta}^2}
{\omega -\varepsilon_{{\bf k}+{\bf Q}}}} \, .
\label{intB}
\end{eqnarray}
The remaining  one dimensional integral  demonstrates the different behavior for  different ${\cal N}$.
For ${\cal N}=1$,   the distribution function of the SDW-gap is centered around zero  whereas
for ${\cal N}=2$ and  $3$ it 
has a maximum for finite $\tilde{\Delta}$.
Performing the saddle point approximation for ${\cal N}=2$ or $3$ yields  SDW-like solutions with
reduced gap $\Delta_o=\sqrt{\frac{1}{2}} \Delta  \approx 0.7071\Delta $ for ${\cal N}=2$ and
$\Delta_o= \sqrt{\frac{2}{3}} \Delta  \approx  0.8165 \Delta$ for ${\cal N}=3$.
However, even for ${\cal N}=1$,  the contribution of the tails of the distribution 
function changes the behavior qualitatively compared to the saddle point approximation
and   solutions similar to an SDW state occur.
It is interesting to  note Eq.~\ref{intB} is the Borel integral
representation of the formally divergent perturbation series in the
limit $\xi \rightarrow \infty$. 
With the path integral  approach, we
did not    encounter 
this divergence, thus 
 demonstrating that it was,
 in fact,
a spurious one.

The integral with respect to the fluctuating SDW gap can be evaluated 
using the integral representation of the incomplete Gamma function
\begin{equation}
\Gamma(\psi, z) =\frac{e^{-z} z^\psi}{\Gamma(1-\psi)} \int_0^\infty \frac{e^{-t} t^{-\psi}}{z+t} \, dt\, ,
\label{gmafkt}
\end{equation}
with $t={\cal N} \tilde{\Delta} ^2/(2 \Delta^2)$ and
$\psi=1-{\cal N}/2$ and $z= -\frac{3}{2}(\omega -\varepsilon_{\bf k})
(\omega -\varepsilon_{{\bf k}+{Q}})/\Delta^2$.   Using   the continued fraction 
representation of $\Gamma(\psi, z)$~\cite{RyGra}, we 
  obtain the following result  for the single particle Greens function:
\begin{equation}
G_{\bf k}(\omega)=\frac{1}{ \displaystyle \omega-\varepsilon_{\bf k} -\frac{\kappa_1 \Delta^2}
{\displaystyle \omega-\varepsilon_{{\bf k}+{\bf Q}} - \frac{  \kappa_2 \Delta^2}
{\displaystyle \omega-\varepsilon_{\bf k} -  \frac{  \kappa_3 \Delta^2}
{\displaystyle \omega-\varepsilon_{{\bf k}+{\bf Q}} -
\cdots }}}}
\label{cfgf}
\end{equation}
with $\kappa_j=j/{\cal N}$ if $j$ even and $\kappa_j=(j+{\cal N}-1)/{\cal N}$ if $j$ odd.
For the special cases of ${\cal N}=1$ and $2$, this is identical to Sadovskii's result~\cite{Sad79}.
Furthermore, we can  obtain an analytical expression  for  the single particle Greens function
 for the case ${\cal N}=3$ which corresponds to the spin fermion model.

\subsubsection{Generalization to the case of  finite $\xi$}
Using, for the moment, the approximation of Eq.~\ref{momapp},
the solution for finite antiferromagnetic correlation length can be inferred
from Eq.~\ref{cfgf}  and 
Eq.~\ref{sigarb}, valid for an arbitrary diagram and finite $\xi $.
 From  Eq.~\ref{sigarb} we know that, compared to the 
limit $\xi \rightarrow \infty$, the only way the correlation length enters
the problem is via:
\begin{equation}
\omega -\varepsilon_{{\bf k}+j{\bf Q}} \rightarrow 
\omega -\varepsilon_{{\bf k}+j{\bf Q}}  +i \,n_j v_{{\bf k}+{\bf Q},j} /\xi \, ,
\label{replace}
\end{equation}
where $j$ refers to the order of the continued fraction with nominator $\kappa_j \Delta^2$.
On the other hand, in Eq.~\ref{sigarb}, the integer number $n_j$ has a specific diagrammatic
meaning.
Since we can generate an arbitrary diagram by expanding the continued fraction of Eq.~\ref{cfgf}
with respect to $\Delta^2$, 
 we can perform the  replacement  of Eq.~\ref{replace} within the continued fraction representation  
Eq.~\ref{cfgf}.
In order to fix the  not yet determined integer number  $n_j$
of the $j$-th insertion of the continued fraction, which is independent  of the 
correlation length, 
we   use the fact that for a given order $\Delta^{2N}$ of the perturbation theory,
 only  one sequence $\{n_j\}$ occurs with $n_j=N$. This refers to diagrams which are identical to
the rainbow diagram of order $2N$.
The only way that this can be  fulfilled  is  $n_j=j$.
Thus we obtain the   solution of the spin fermion model for finite correlation length
based on the approximation of Eq.~\ref{momapp}:
\begin{equation}
{G^{(j)}_{\bf k}(\omega)}^{-1}=  g^{(j)}_{\bf k}(\omega)^{-1}-
\kappa_{j+1} \Delta^2 G^{(j+1)}_{\bf k}(\omega).
\label{sadselfapp}
\end{equation}
with 
 \begin{equation}
{g^{(j)}_{\bf k}(\omega)}=\frac{1}{\omega -
\varepsilon_{{\bf k}+j{\bf Q}}+i \frac{j v_{{\bf k},j}}{\xi} }
\end{equation}
which generates the continued fraction representation of the single particle Green's function,
similar to the one dimensional case.

The general  solution, independent of Eq.~\ref{momapp}, 
follows from the fact that the only difference  from  the strict  two dimensional case
is the   function $\psi_{{\bf k}+{\bf Q}}(t)$ of Eq.~\ref{psidef}, which was approximated by
Eq.~\ref{psi2}.
The decoupling of the momentum integrals in the proper time representation and
the diagram multiplicities, i.e. the $\kappa_j$, do not depend on the actual choice of  
$\psi_{{\bf k}+{\bf Q}}(t)$, and  the only difference compared to Eq.~\ref{sadselfapp}
is the function $g^{(j)}_{\bf k}(\omega)$, which in terms of $\psi_{{\bf k}+{\bf Q}}(t)$
can be expressed as
\begin{equation}
g^{(j)}_{\bf k}(\omega) =-i \int_0^\infty dt \, 
e^{i(\omega-\varepsilon_{{\bf k}+j{\bf Q}})}\,  \psi_{{\bf k}+j{\bf Q}}(t)^j\, .
\end{equation}
On using the result of Eq.~\ref{psi1} for $\psi_{{\bf k}+{\bf Q}}(t)$,
together
with Eq.~\ref{sadselfapp},
 the solution of the Green's function of the spin fermion model
is  that given in Eq.~\ref{sadself}.



\end{multicols}
%
%
%
%
%
%
%
%
\newpage
\begin{figure}
\centerline{\epsfig{file=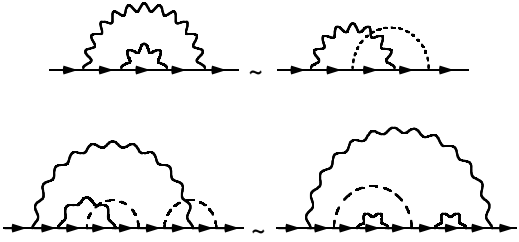,width=9cm, height=7cm}}
\caption{Self energy diagrams of order $g^4$  and $g^8$, which are,
 within the quasistatic approximation, identical apart from 
multiplicity
 and sign, because the number of  spin fluctuation lines (shown by 
wiggly lines for
 transverse
and dashed lines for longitudinal  spin excitations) on top
of a given electron propagator (solid line) are the same.  }
\label{fT2}
\end{figure}
 
\begin{figure}
\centerline{\epsfig{file=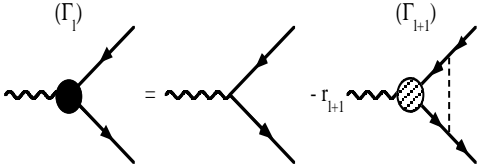,width=8cm, height=4cm, scale=2}}
 \caption{Diagrammatic illustration of the recursion relation of the
  vertex function, which is similar to the ladder approximation of the
  irreducible vertex.  All non-ladder diagrams are taken into account
  by the corresponding weighting factors $r_{l+1}$.}
\label{fT3}
\end{figure}

\begin{figure}
\centerline{\epsfig{file=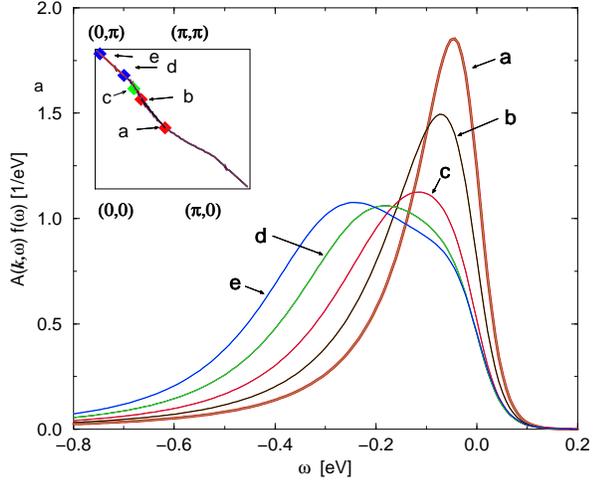,width=9cm, height=7.38cm, scale=1}}
\caption{ The spectral density multiplied with Fermi function on the 
 Fermi surface  for  $\xi=3$.The distinct behavior of  hot and 
cold quasiparticles is visible.  }
\label{fR3}
\end{figure}
%
\begin{figure}
\centerline{\epsfig{file=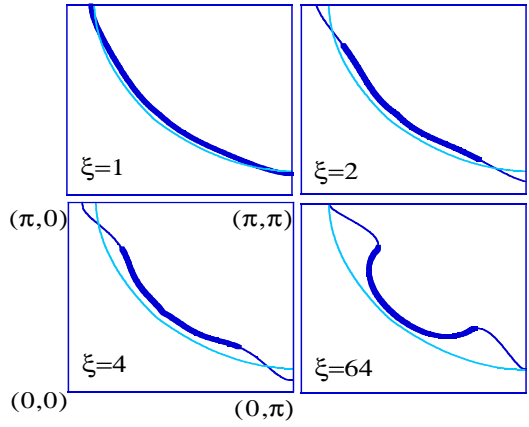,width=8cm, height=6.56cm, scale=1}}
\caption{ Fermi surface (solid line), in comparison with the bare
Fermi surface (dotted line) and the 
visible part of the Fermi surface (thick solid line), i.e 
  only for momenta where a maximum of the spectral density crosses
the Fermi energy  if ${\bf k}$ crosses the Fermi surface.  }
\label{fR4}
\end{figure}

\begin{figure}
\centerline{\epsfig{file=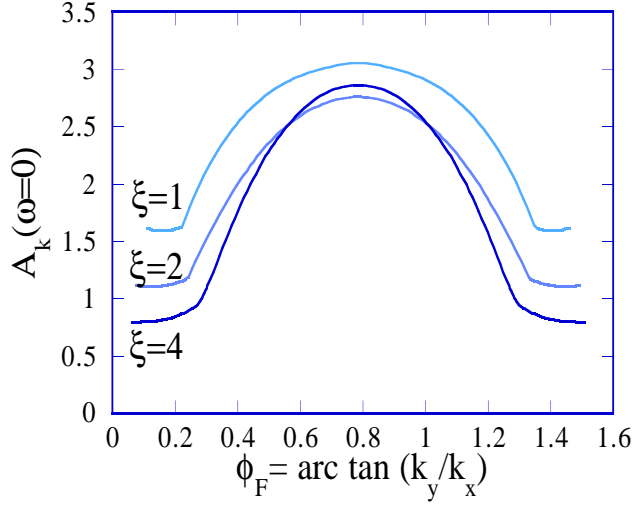,width=10cm, height=8cm, scale=1}}
 \caption{ The spectral density $A_{\bf k}(\omega=0)$ along the Fermi surface 
as function of $\phi_{\rm F}={\rm arctan}(k_y/k_x)$.  }
\label{fR5}
\end{figure}

\begin{figure}
\centerline{\epsfig{file=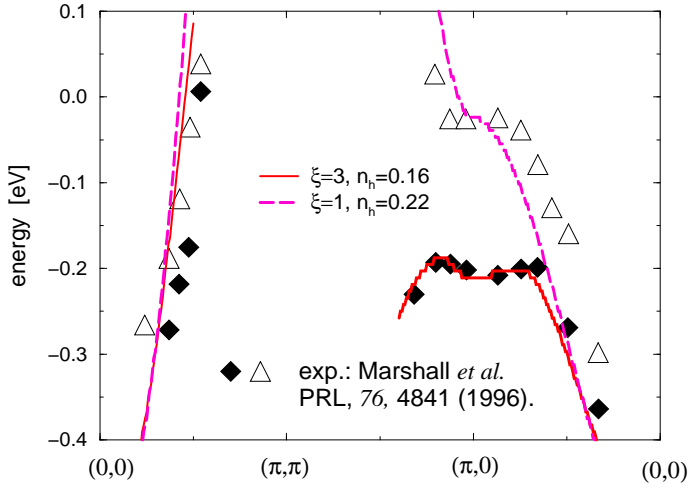,width=10cm, height=8.2cm, scale=1}}
\caption{the momentum dependence of local maxima of the spectral density 
as function of 
  $\xi$
and hole doping concentration $n_h$ is compared with the 
experiments of Ref.\ [25]
for Bi$_2$Sr$_2$Ca$_{1-x}$Dy$_x$Cu$_2$O$_{8+\delta}$ with $x=0.1$ (triangles)
and $x=0.175$ (diamonds).
Only maxima with relative spectral weight $> 10\%$ are shown.     }
\label{fR6}
\end{figure}

\begin{figure}
\centerline{\epsfig{file=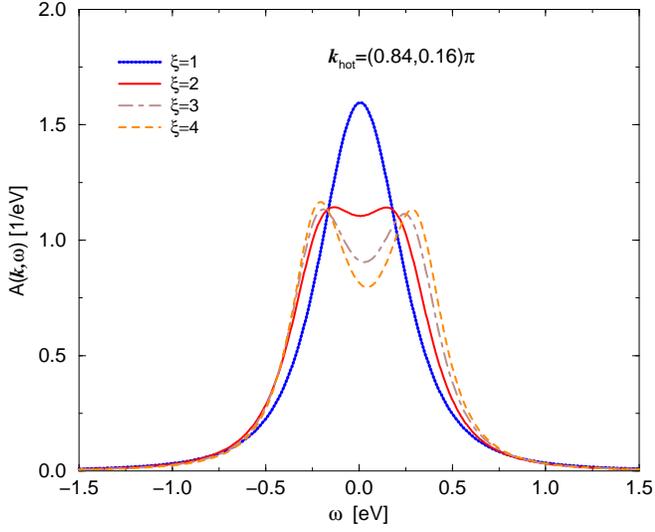,width=10cm, height=8.2cm, scale=1}}
\caption{  $\xi$ dependence of the spectral density at the hot spot. The appearance
of SDW precursors for $\xi > 2$ can be seen. For smaller values of $\xi$ 
the system behaves
conventionally. }
\label{fR7}
\end{figure}

\begin{figure}
\centerline{\epsfig{file=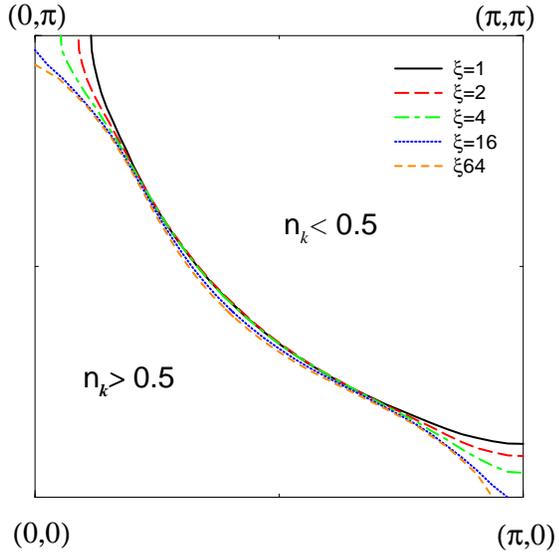,width=10cm, height=8.2cm, scale=1}}
\caption{ ${\bf k}$ -points with $n_{\bf k}=\frac{1}{2}$ for different
correlation lengths in comparison with the results for an
uncorrelated Fermi system. }
\label{fR2}
\end{figure}
\begin{figure}
\centerline{\epsfig{file=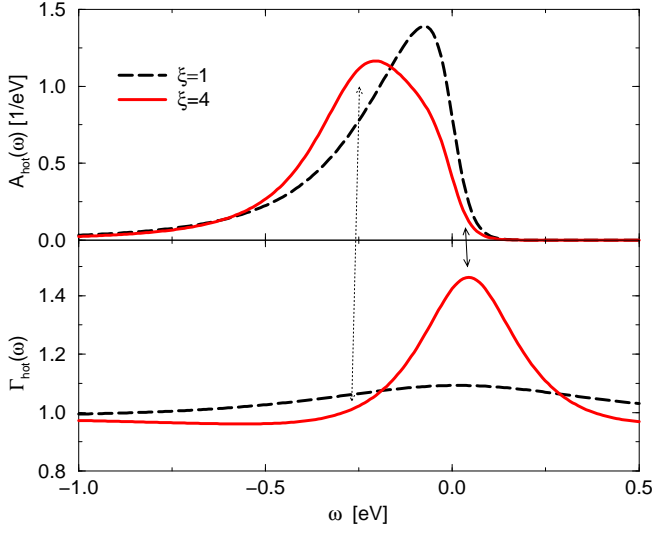,width=10cm, height=8.2cm, scale=1}}
\caption{ The   spectral density $A_{\bf k}(\omega)$ and 
spin vertex  function $\Gamma^{ {\rm s}}_{{\bf k},{\bf k}+{\bf Q}}
(\omega+i0^+,\omega+i0^+) $ for a hot   quasiparticle with momentum transfer 
${\bf Q}$ and zero frequency transfer as function of energy  are shown for
two correlation lengths.}
\label{fR8}
\end{figure}

\begin{figure}
\centerline{\epsfig{file=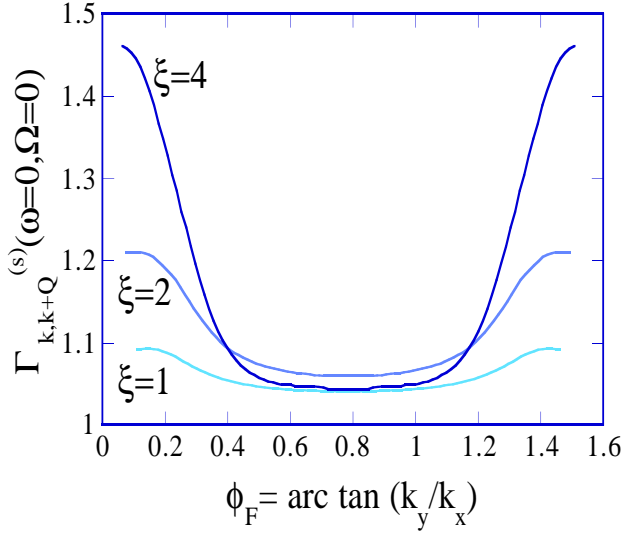,width=10cm, height=8.2cm, scale=1}}
\caption{Spin vertex $ \Gamma^{ {\rm s}}_{{\bf k},{\bf k}+{\bf Q}}
(\omega+i0^+,\omega+i0^+) $
 for $\omega=0$ along the Fermi surface 
as function of $\phi_{\rm F}={\rm arctan}(k_y/k_x)$.   }
\label{fR9}
\end{figure}

\begin{figure}
\centerline{\epsfig{file=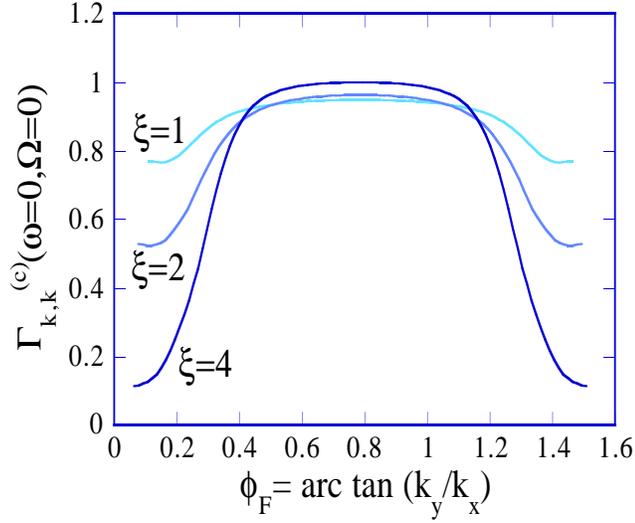,width=10cm, height=8.2cm, scale=1}}
\caption{Charge vertex  
$ \Gamma^{ {\rm c}}_{{\bf k},{\bf k}+{\bf Q}}
(\omega+i0^+,\omega+i0^+) $  for $\omega=0$  as a function of  
$\phi_{\rm F}={\rm arctan}(k_y/k_x)$. }
\label{fR10}
\end{figure}
  
\begin{figure}
\centerline{\epsfig{file=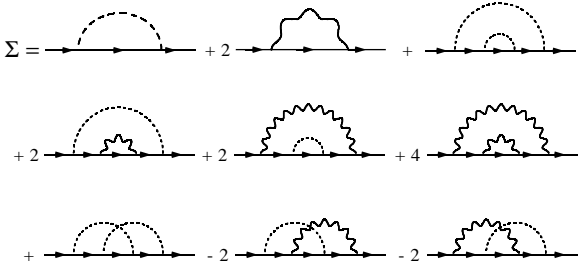,width=9cm, height=7.5cm, scale=2}}
\vskip -2cm
\caption{Self energy diagrams of the spin fermion model up to order
$g^4$, including signs and multiplicities. The wiggly (dashed) lines
correspond to spin flip (spin conserving) processes.
Each  crossing of a wiggly and dashed lines causes a prefactor $-1$ and each
wiggly line an additional factor of $2$. The solid line corresponds to
the bare  fermion propagator and each vertex to the coupling 
constant $g/ 3^{1/2}$.} 
\label{fAp1}
\end{figure}

\begin{figure}
   
\centerline{\epsfig{file=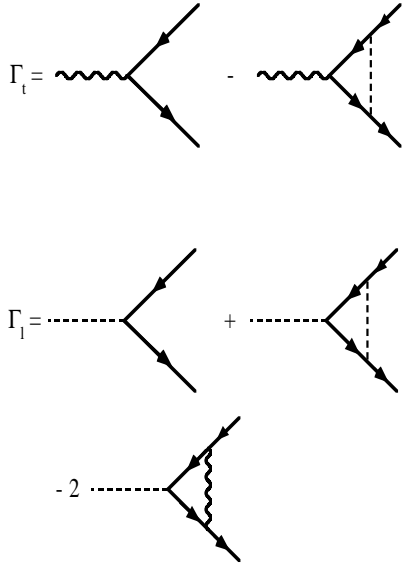,width=13cm, height=19cm, scale=2}}
\vskip -9cm
\caption{  Longitudinal (spin conserving) and transverse (spin flip) 
vertices  up to $g^2$. Note, that the derived diagrammatic rules
guarantee, as expected,
spin rotation invariance: $\Gamma_{\rm l}= \Gamma_{\rm t} $. }
\label{fAp2}
\end{figure}
 
\end{document}